\documentclass[twocolumn]{aastex63}

\usepackage{color}
\usepackage{graphicx,graphics}
\usepackage{amssymb,amsmath}
\usepackage{natbib}
\usepackage {threeparttable}
\usepackage{bm}

\newcommand{\Mstar}{M_\star}
\newcommand{\HI}{{\sc Hi}}
\newcommand{\Msun}{{\rm  M_{\odot}}}

\shorttitle{Connection Between Galaxies and {\sc Hi} in the CGM and IGM}
\shortauthors{R. Momose et al.}

\begin{document}

\title{
Connection Between Galaxies and {\sc Hi} in the Circumgalactic and Intergalactic Media: \\ 
Variation According to Galaxy Stellar Mass and Star-formation Activity
}


\correspondingauthor{Rieko Momose}
\email{momose@astron.s.u-tokyo.ac.jp}

\author[0000-0002-8857-2905]{Rieko Momose}
\affiliation{Department of Astronomy, School of Science, The University of Tokyo, 7-3-1 Hongo, Bunkyo-ku, Tokyo 113-0033, Japan}

\author{Ikkoh Shimizu}
\affil{Shikoku Gakuin University, 3-2-1 Bunkyocho, Zentsuji, Kagawa 765-0013, Japan}
\affil{National Astronomical Observatory of Japan, 2-21-1 Osawa, Mitaka, Tokyo 181-8588, Japan}

\author[0000-0001-7457-8487]{Kentaro Nagamine}
\affil{Theoretical Astrophysics, Department of Earth and Space Science, Graduate School of Science, Osaka University, \\
1-1 Machikaneyama, Toyonaka, Osaka 560-0043, Japan}
\affil{Kavli-IPMU (WPI), The University of Tokyo, 5-1-5 Kashiwanoha, Kashiwa, Chiba 277-8583, Japan}
\affil{Department of Physics \& Astronomy, University of Nevada, Las Vegas, 4505 S. Maryland Pkwy, Las Vegas, NV 89154-4002, USA}

\author[0000-0002-2597-2231]{Kazuhiro Shimasaku}
\affiliation{Department of Astronomy, School of Science, The University of Tokyo, 
7-3-1 Hongo, Bunkyo-ku, Tokyo 113-0033, Japan}
\affiliation{Research Center for the Early Universe, The University of Tokyo, 7-3-1 Hongo, Bunkyo-ku, Tokyo 113-0033, Japan}

\author[0000-0003-3954-4219]{Nobunari Kashikawa}
\affiliation{Department of Astronomy, School of Science, The University of Tokyo, 
7-3-1 Hongo, Bunkyo-ku, Tokyo 113-0033, Japan}
\affiliation{Research Center for the Early Universe, The University of Tokyo, 7-3-1 Hongo, Bunkyo-ku, Tokyo 113-0033, Japan}

\author[0000-0002-3801-434X]{Haruka Kusakabe}
\affiliation{Observatoire de Gen\`{e}ve, Universit\'e de Gen\`{e}ve, 51 chemin de P\'egase, 1290 Versoix, Switzerland}

\begin{abstract}

This paper systematically investigates comoving~Mpc scale intergalactic medium (IGM) environment around galaxies traced by the Ly$\alpha$ forest. Using our cosmological hydrodynamic simulations, we investigate the IGM--galaxy connection at $z=2$ by two methods:  
(I) cross-correlation analysis between galaxies and the fluctuation of Ly$\alpha$ forest transmission ($\delta_\text{F}$); and 
(II) comparing the overdensity of neutral hydrogen ({\sc Hi}) and galaxies. 
Our simulations reproduce observed cross-correlation functions (CCF) between Ly$\alpha$ forest and Lyman-break galaxies. 
We further investigate the variation of the CCF using subsamples divided by dark matter halo mass ($M_\text{DH}$), galaxy stellar mass ($\Mstar$), and star-formation rate (SFR), and find that the CCF signal becomes stronger with increasing $M_\text{DH}$, $\Mstar$, and SFR.
The CCFs between galaxies and gas-density fluctuation are also found to have similar trends.
Therefore, the variation of the $\delta_\text{F}$--CCF depending on $M_\text{DH}$, $\Mstar$, and SFR is due to varying gas density around galaxies. 
We find that the correlation between galaxies and the IGM {\sc Hi} distribution strongly depends on $M_\text{DH}$ as expected from the linear theory.
Our results support the $\Lambda$CDM paradigm, finding a spatial correlation between galaxies and IGM {\sc Hi}, with more massive galaxies being clustered in higher-density regions.

\end{abstract}


\section{introduction}

The standard picture of galaxy formation within the gravitational instability paradigm indicates that galaxy formation and evolution is closely linked to its surrounding gas called the circumgalactic medium (CGM) and intergalactic medium (IGM) (e.g., \citealp{rauch98,mo10}).
The inflowing gas from the IGM provides the fuel for star formation in galaxies, and promotes the growth of galaxies and their central supermassive black hole (SMBH). 
As important as the inflow is the energetic feedback from  massive stars and SMBHs which blows the gas away into the CGM and IGM.  
Therefore, determining the Mpc-scale distribution of gas as a function of time and space is quite important for understanding galaxy formation and evolution. 

The connection between the CGM/IGM and galaxies has been studied using Ly$\alpha$ forest absorptions in quasar spectra.  
The most common method to clarify the CGM/IGM--galaxy connection is the cross-correlation analysis between  Ly$\alpha$ forest absorption and galaxies (e.g., \citealp{adelberger03,adelberger05,chen05,ryan-weber06,FG08,rakic11,rakic12,rudie12,font-ribera13,profX13,tejos14,bielby17}). 
Alternatively, comparisons of galaxy and neutral hydrogen ({\sc Hi}) overdensities have also been discussed in the literature (\citealp{mukae17,mukae19,mawatari17}). 
Specific high-density regions with abundant {\HI} gas and highly clustered galaxies have also been studied  (e.g., \citealp{cai16,lee16,mawatari17,hayashino19}).
All of the above studies have revealed that galaxy distribution correlates with the IGM up to tens of comoving Mpc scales. 

Physical properties of {\HI} gas in the CGM or IGM have been studied in detail theoretically
(e.g., \citealp{meiksin09,meiksin14,meiksin17,fumahalli11,vandeVoort12a,vandeVoort12b,rahmati15}), aided by powerful cosmological hydrodynamic simulations such as EAGLE \citep{schaye15}, Illustris \citep{Vogelsberger14a,Vogelsberger14b,genel14,sijacki15}, and IllustrisTNG \citep{Villaescusa-Navarro18,nelson19}. 
\citet{turner17} presented the median {\HI} optical depth vs. line-of-sight or transverse distance around galaxies in the EAGLE simulation, and found that it is sensitive to dark matter halo mass. 
\citet{sorini18} compared the mean Ly$\alpha$ absorption profile as a function of the transverse distance from galaxies in both observations and simulations, and have shown a reasonable match between them beyond $2$ proper Mpc (pMpc), but significant differences at $0.02-2$\,pMpc.
The subsequent study by \citet{sorini20} has investigated the impact of feedback around mock quasars and found that the stellar feedback is the more dominant driver (rather than AGN, active galactic nuclei) to determine the average properties of the CGM.

The correlation between galaxies and IGM morphology has also been examined in the literature. 
\citet{martizzi19} have demonstrated that galaxies with lower stellar masses than the median are in voids and sheets of the IGM, whereas galaxies with higher stellar masses are more likely to be in filaments and knots of the IGM with higher gas densities. 
In addition, the correlation of mock Ly$\alpha$ forest absorption spectra with galaxy overdensity has been studied and compared with observations (e.g., \citealp{stark15,miller19}). 
Although these theoretical studies provide further evidence of a strong link between the CGM/IGM and galaxies, there are many aspects that are still unclear and our understanding is still insufficient. 

Why do we need to study the distribution of {\sc Hi} and its bias beyond the well-established linear theory of the $\Lambda$CDM model? One of the main reasons is that the bias between baryons, galaxies, halos, and dark matter is non-trivial, and it is of substantial interest for future projects of IGM tomography, such as the Subaru PFS, Euclid, DESI, and 21cm cosmology by SKA. Understanding nonlinear bias at intermediate and small scales is becoming increasingly more important for the BAO studies of precision cosmology.  
What we are trying to achieve in this paper is to dig deeper into this nonlinear bias between baryons (particularly {\sc Hi}), galaxies, and dark matter, beyond the simple halo bias, and evaluate the impact of limited observational data on the cross-correlation function (CCF). Feedback effects from galaxies can alter the {\sc Hi} distribution in the intergalactic space due to ionization, and it is crucial to use a cosmological hydrodynamic simulation that solves heating and cooling self-consistently with the impact of the UV background radiation field. As observations can only trace a certain phase of gas, it is unclear if the simple bias of linear theory is applicable to {\sc Hi}.  In fact, it has been shown that {\sc Hi} has a complicated scale-dependent bias by many authors (e.g., \citealp{Villaescusa-Navarro18,ando19,ando20,Sinigaglia20}).
It is thus crucial to examine the IGM {\sc Hi}--galaxy connection more deeply, including the dependence on galaxy mass and galaxy population. 
Current observational data is still very limited to study the CCF, both in terms of the survey volume of Ly$\alpha$ forest absorptions and the number of galaxies with spectroscopic redshift (\citealp{momose20b}; hereafter M21).
In comparison, numerical simulations provide an alternative opportunity to study the IGM {\sc Hi}--galaxy connection, and this paper presents such systematic investigations for the first time.  Our results certainly include the effect of halo bias, but they also reflect the intricate nonlinear bias between {\sc Hi}, galaxies, and dark matter, beyond the simple linear theory.

In order to unveil the gas distribution on Mpc-scales around galaxies, we systematically investigate it using numerical simulations. We aim to: 
1) establish the methodology to evaluate the CGM/IGM--galaxy connection; 
2) examine its dependency on galaxy properties; and 
3) verify the cause of dependencies using {\small GADGET3-Osaka} cosmological hydrodynamic simulations (\citealp{shimizu19}). 
What is new in this paper is a systematic investigation of the IGM--galaxy connection depending on several galactic properties. We make subsamples based on galaxies' stellar mass $\Mstar$, halo mass $M_\text{DH}$, star-formation rate (SFR) and specific SFR (sSFR = SFR/$\Mstar$), and compare the CCFs.
Calculating the CCFs in three-dimensions taking various galaxy properties into account is also a novel aspect of this work. Studying the IGM--galaxy correlation with the same resolution as the latest observations and comparing the simulation results with them is clearly a worthwhile exercise to check the validity of the $\Lambda$CDM model because the application of linear theory to the IGM {\sc Hi}--galaxy connection has not yet been fully investigated. 
In this study, we particularly focus on the statistical comparisons between simulations and observations. 
For a fair comparison, we use the same parameters as in our companion observational paper (M21).
As we describe in more detail in M21, for observations, we use the CLAMATO (COSMOS Ly$\alpha$ Mapping And Tomography Observations) which is publicly-available Ly$\alpha$ forest 3D tomography data \citep{lee14,lee16,lee18}, and several other catalogs in the archives.

This paper is organized as follows. 
We introduce our numerical simulations in Section 2 and the methodology to examine the IGM--galaxy connection in Section 3.
Results and discussion are presented in Sections 4 and 5, respectively. 
Finally, we give our summary in Section 6. 
In the appendix, we discuss the dependencies of our results on redshift width, redshift uncertainty, sample size, and cosmic variance. 
We note that ``cosmic web'' and ``IGM'' are used specifically for those traced by {\HI} gas unless otherwise specified in this paper. 
In addition, we mainly use $h^{-1}$ Mpc in comoving units in the following sections.

\section{Simulations}
\label{sec:sim}

In this paper, we use the cosmological hydrodynamic simulations performed with {\small GADGET3-Osaka} \citep{shimizu19}, which is a modified version of the Tree-PM SPH code {\scriptsize GADGET-3} \citep[originally described in][]{Springel2005}. 
Some physical processes important for galaxy formation such as star formation, supernova (SN) feedback and chemical enrichment have been implemented and described in detail by \citet{shimizu19}. 
Our simulations reproduce various observational results such as stellar mass function, SFR function, stellar-to-halo-mass ratio, and cosmic star formation history within observational uncertainties at $z \geq 2$ (Shimizu et al. 2020, in preparation).

Here, we briefly describe our simulations, which employ
$N=2 \times 512^3$ particles in a comoving volume of ($100\,h^{-1}$\,${\rm Mpc})^3$. 
The particle masses of dark matter and gas are $5.38 \times 10^8\,\Msun$ and $1.00 \times 10^8\,h^{-1}\,\Msun$, respectively. 
The gravitational softening length is set to be $8\,h^{-1}$\,${\rm kpc}$ in comoving units. 
For SPH particles, the smoothing length is allowed to be 10\% of the softening length ($800\,h^{-1}$\,${\rm pc}$ in comoving units) in regions where baryons dominate over dark matter. 
This means that at $z=3$ the physical maximum resolution for gas may reach 200\,pc (proper) in our simulations. 
In this study, we pay particular attention to the physical properties of gas at $r>100\,h^{-1}$\,${\rm kpc}$ from galaxies, and our resolution is sufficient for this study.

Star particles are generated from gas particles when a set of criteria are satisfied. 
Note that the mass of gas particles changes over time due to star formation and stellar feedback (by supernova and AGB stars). 
As described in detail in \citet{shimizu19}, we employ the CELib package \citep{saito17} which can treat the time and metallicity-dependent metal yields and energies from SNII, SNIa and AGB, allowing for more realistic chemical evolution of simulated galaxies. 
In order to identify the simulated galaxies, 
we run a friends-of-friends (FoF) group finder with a comoving linking length of $0.2$ in units of the mean particle separation to identify groups of dark matter particles as dark matter halos. 
We then identify gravitationally-bound groups of minimum 32 particles (dark matter + SPH + star) 
as substructures (subhalos) in each FoF group using the {\scriptsize SUBFIND} algorithm \citep{Springel2001}, which is a standard practice in many of the simulation works. 
We regard substructures that contain at least five star particles as our simulated galaxies. 
Moreover, we define the most massive galaxy in a halo as the central galaxy, and the rest as satellite galaxies. 
We also calculate the virial halo mass $(M_{\rm DH})$, which is defined by the total enclosed mass inside a sphere of 200 times the critical density of the Universe.  
This means that the member galaxies (central and satellite galaxies) in a dark matter halo have the same $M_{\rm DH}$, even though the substructures can have different subhalo masses calculated by {\scriptsize SUBFIND}. 

Note that each gas (star) particle has some associated physical properties such as mass, SFR and metallicity. 
In our SPH simulation, these values are automatically computed following hydrodynamic and gravitational interactions, rather than given by hand, which is a fundamental difference from the semi-analytic models of galaxy formation.  
The properties (gas mass, stellar mass $\Mstar$, and SFR) of a simulated galaxy are defined by summation of these quantities in each subhalo. 

In order to directly compare our simulations and observations,  we create the light-cone output of gas particles and galaxies by connecting $10$ simulation boxes of different redshifts 
following our previous work \citep{Shimizu2012, Shimizu2014, Shimizu2016}. 
The redshift range of our light-cone output is from $z \sim 1.8$ to $3.1$ which can cover the redshift range of recent Ly$\alpha$ absorption line surveys \citep[e.g., CLAMATO;][]{lee14, lee16, lee18} and future PFS Ly$\alpha$ absorption survey \citep{Takada2014}. 
We then randomly shift and rotate each simulation box so that the same objects do not appear multiple times on a single line-of-sight (LoS) at different epochs. 

With this light-cone output, we calculate the Ly$\alpha$ optical depth ($\tau_{\rm Ly \alpha}$) along the LoS. 
First, we calculate the important physical quantities, $A_{\rm grid}(x)$, at each grid point $x$ along LoS, such as {\HI}  density, LoS velocity and temperature as follows: 
\begin{equation}
A_{\rm grid}(x) = \sum_j \frac{m_j}{\rho_j} A_j W(r, h_j), 
\end{equation}
where $A_j$, $m_j$, $\rho_j$ and $h_j$ are the physical quantity of concern, gas particle mass, gas density, and smoothing length of $j$-th particle, respectively. 
$W$ is the SPH kernel function, and $r$ is the distance between LoS grid points and gas particles.
For simplicity, the grid size ($dl$) is set to a constant value of $100\,h^{-1}$\,kpc in comoving units which corresponds to 10\,${\rm km\,s^{-1}}$ in the velocity space at $z\simeq 2$. This is a higher resolution than any of the relevant Ly$\alpha$ observations. 
Then, we calculate the Ly$\alpha$ optical depth $\tau_{\rm Ly\alpha}(x)$ using these physical values at each grid point as follows:  
\begin{equation}
\tau_{\rm Ly\alpha}(x) = \frac{\pi e^2}{m_e c}  \sum_j f_{ij}\, \phi(x - x_j)\, n_{\rm HI}(x_j) dl,  
\end{equation}
where $e$, $m_e$, $c$, $f_{ij}$, $n_{\rm HI}$, and $x_j$ are the electron charge, electron mass, speed of light, absorption oscillator strength, {\HI} number density, and $j$-th grid point location, respectively. 
$\phi$ is the Voigt profile, and we use the fitting formula of \citet{Tasitsiomi2006} without direct integration. 
In this study, after making the high resolution LoS data, we reduce our resolution by coarse-graining the grid size to match the observations. 
1024 $(=32^2)$ LoSs are drawn with regularly spaced intervals. 
The mean separation of each LoS is 3.3 $h^{-1}$ Mpc which is similar to the CLAMATO survey. 

Finally, we note that the AGN feedback is not considered in the simulation that we use for this paper. 
We note that the impact of AGN feedback on {\HI} distribution is still under debate. 
For example, \citet{sorini20} have found little impact of AGN feedback on the surrounding CGM in general, and that SN feedback has more dominant effects. Besides, statistical results of the {\sc Hi}--galaxy connection obtained from the same simulations as this study have shown good agreement with current observations \citep{Nagamine20}. 
Some observations have found the proximity effect around quasars (e.g., \citealp{mukae19,momose20b}), which requires full radiative transfer calculations to compare with simulations. 
Therefore, we defer the study of AGN feedback using radiation transfer to our future work.


\begin{deluxetable*}{ccrl}
\tablenum{1}
\label{tab:num}
    \tablecaption{The Number of Galaxies in Each Sample}
\tablehead{
    \colhead{Category} & \colhead{Range} & \colhead{Number} & \colhead{Sample Name} 
}
\startdata
    All galaxies & & 89446 &     \\
\hline
    Stellar mass $\Mstar$ [$h^{-1}$ M$_\sun$]    & $10^{11}\leq \Mstar$           & 1662     & $\Mstar$--$11$ \\
     & $10^{10}\leq \Mstar < 10^{11}$ & 21975    & $\Mstar$--$10$\\
    & $10^{9}\leq \Mstar < 10^{10}$  & 65809    & $\Mstar$--$9$\\
\hline
    Halo mass $M_\text{DH}$ [$h^{-1}$ M$_\sun$]  & $10^{13}\leq M_\text{DH}$           & 1874   & $M_\text{DH}$--$13$ \\
    & $10^{12}\leq M_\text{DH} < 10^{13}$ & 20407  & $M_\text{DH}$--$12$ \\
    & $10^{11}\leq M_\text{DH} < 10^{12}$ & 66803  & $M_\text{DH}$--$11$ \\
    & $10^{10}\leq M_\text{DH} < 10^{11}$ & 362    & $M_\text{DH}$--$10$ \\
\hline
    $\log$ (SFR/M$_\sun$ yr$^{-1}$)      & $2\leq \log$ SFR       & 1152      & SFR--(i) \\
    & $1\leq \log$ SFR $<2$  & 24349     & SFR--(ii) \\
    & $0\leq \log$ SFR $<1$  & 46425     & SFR--(iii)\\
    & $-1\leq \log$ SFR $<0$ & 14654     & SFR--(iv)\\
    & $-1>\log$ SFR          & 2866      & SFR--(v)\\
\hline
    $\log$ (sSFR/yr$^{-1}$)     & $-9\leq \log$ sSFR        & 38078 & sSFR--(i) \\
    & $-10\leq \log$ sSFR $<-9$ & 47419 & sSFR--(ii) \\
    & $-10> \log$ sSFR          & 3949  & sSFR--(iii) \\
\enddata
\end{deluxetable*}
\label{tab:num}


\section{methodology}
\label{sec:method}

One of the main purposes of this study is to compare with the CLAMATO, which is a 3D tomography data of Ly$\alpha$ forest transmission fluctuation ($\delta_\text{F}$: see the following definition) over $2.05<z<2.55$ in $0.157$ deg$^2$ of the COSMOS field \citep{scovil07,lee16,lee18}. 
The CLAMATO consists of $60\times48\times876$ pixels corresponding to $30\times 24\times436$ $h^{-1}$ Mpc cubic with a pixel size of $0.5$ $h^{-1}$ Mpc. 
The average separation of background galaxies for measuring Ly$\alpha$ absorption are [$2.61$, $3.18$]\,$h^{-1}$\,Mpc in [RA, DEC] directions, and the separation in LoS-direction is $2.35 h^{-1}$\,Mpc at $z\sim2.3$. (See M21 for more detailed comparison with CLAMATO.)

To produce a similar data cube of $\delta_\text{F}$ from our LoS data, we first evaluate the Ly$\alpha$ transmission fluctuation $\delta_\text{F}$ in each LoS pixel by
\begin{equation}
    \delta_\text{F} \equiv \frac{F(x)}{\langle F_z(x) \rangle} - 1,
\end{equation}
where $F(x)= \exp{[-\tau_{\rm Ly\alpha}(x)]}$ is the Ly$\alpha$ flux transmission, and $\langle F_z(x) \rangle$ is the cosmic mean transmission.
We adopt the following value derived by \citet{FG08}: 
\begin{equation} 
    \langle F_z(x) \rangle = \exp{[-0.00185(1+z)^{3.92}]},
\end{equation}
because it is used in CLAMATO. 
Additionally, we also use the same setup used in CLAMATO with Hubble constant $h=0.7$ and redshift coverage of $2.05\leq\,z\,\leq2.55$.


\vspace{0.5cm}

In the following two subsections, we present the methods for two analyses: (I) cross-correlation, and (II) overdensity analysis. 

\subsection{Cross-correlation Analysis}

The cross-correlation analysis 
is often used in the literature to characterize the correlation between galaxies and CGM/IGM (e.g., \citealp{adelberger05,tejos14,croft16,croft18,bielby17}). 
In this study, we adopt the following definition of CCF: 
\begin{equation}
  \xi_\text{$\delta$F}(r) = \frac{1}{N(r)} \sum_{i=1}^{N(r)} \delta_{g, i}- \frac{1}{M(r)} \sum_{j=1}^{M(r)} \delta_{ran, j}, 
  \label{eq:cross-corr}
\end{equation}
where $\xi_\text{$\delta$F}$ is the CCF between $\delta$F and galaxies; 
$\delta_{g, i}$ and $\delta_{ran, j}$ are the values of $\delta_\text{F}$
for the pixel $i$ and $j$ at the distance $r$ from galaxies and random points \citep{croft16}.
$N(r)$ and $M(r)$ are the numbers of pixel-galaxy and pixel-random pairs in the bin with the distance $r$.

To calculate CCFs, we prepare two LoS data with different LoS grid resolution.
One is the LoS data with original resolution of comoving $0.1$ $h^{-1}$\,Mpc. 
The other is a lower resolution data with coarse-grained grid size of $0.4$\,$h^{-1}$\,Mpc (as described in Section~\ref{sec:sim}) to match the CLAMATO resolution of $0.5$\,$h^{-1}$\,Mpc at $z=2.35$.
We call this latter lower-resolution dataset as `LoS-4'.
For comparison with observations, we use the LoS-4 dataset and calculate CCFs at $1 - 50$\,$h^{-1}$\,Mpc scale around galaxies. 

At the same time, we also use the original LoS dataset to derive CCFs at $r=0.16-1$\,$h^{-1}$\,Mpc, because the redshift resolution of LoS-4 data is larger than the smallest radius for CCF calculation. 
Hereafter, we refer to the scales of $r<1$\,$h^{-1}$\,Mpc and $r\geq1$\,$h^{-1}$\,Mpc as the `CGM regime' and `IGM regime' as indicated in Figure \ref{fig:IM_all_CCF}--1(e). 

The $\xi_\text{$\delta$F}$ value is evaluated in each shell with a thickness of $\Delta \log (r/h^{-1}$ Mpc) $=0.2$ and 0.1 for the CGM and IGM regimes, respectively. 
We confirm that our CCFs by LoS and LoS-4 data are smoothly connected at $r=1$\,$h^{-1}$\,Mpc within the error. 
We perform Jackknife resampling by leaving one object out and calculating $\xi_\text{$\delta$F}$ value, and adopt Jackknife standard error as the error in each shell.

To examine how the CCF varies according to the physical properties of galaxies,
We divide the galaxy sample into $3-5$ subsamples according to $\Mstar$, $M_\text{DH}$, SFR and sSFR. 
The number of galaxies in each subsample and its name are summarized in Table~\ref{tab:num}.

\subsection{Overdensity Analysis}

Another analysis that we perform in this paper is a direct comparison of the IGM absorption and galaxy overdensity within cylinders along the LoS direction. 
This method was originally proposed by \citet{mukae17}, and 
we call it the ``Overdensity analysis''. 
We first generate a 2D LoS map by binning the LoS data with $\Delta z=0.032$, which corresponds to $27.9\,h^{-1}$\,Mpc. 
We then estimate the mean IGM fluctuation $\langle \delta_\text{F} \rangle$ within circles of radius $4.74$\,$h^{-1}$\,Mpc centered on local minima and maxima of the map. 
The sizes of both $\Delta z$ and cylinder radius are chosen to be comparable to the actual observations (see M21).
Note that we use a cylinder of $\Delta z=0.08$ ($\Delta z=0.032$) corresponding to $69.7$ ($27.9$)\,$h^{-1}$\,Mpc in length, and with a radius of $3\,(4.74)\,h^{-1}$\,Mpc in our observational analysis because the mean separation of our LoS data is $2.35\,h^{-1}$\,Mpc.
To determine the pixel positions of local min/max,  
we first mark the positions of local min/max of $\delta_\text{F}$ within a few $h^{-1}$\,Mpc scale, and then further repeat the same procedure to identify the min/max on even smaller scales. 

The galaxy overdensity is also computed together with $\langle \delta_\text{F} \rangle$ in the same cylinders as follows:
\begin{equation}
    \Sigma_\text{gal} = \frac{N_\text{gal}}{\langle N_\text{gal} \rangle} - 1,
\end{equation}
where $N_\text{gal}$ and $\langle N_\text{gal} \rangle$ are the exact number of galaxies and the mean number of galaxies in the cylinder, respectively.
Note that we calculate $N_\text{gal}$ and $\langle N_\text{gal} \rangle$ for each of the three stellar-mass divided subsamples given in Table~\ref{tab:num}.
Since we first generate a 2D LoS map, the galaxy overdensity computed above can be regarded as galaxy surface density, and thus we denote it as $\Sigma_\text{gal}$.

To increase the number of data points, we randomly select four different redshift slices. The number of galaxies in each redshift slice is summarized in Table \ref{tab:overdensity}.
We also perform the overdensity analysis for randomly selected positions to examine possible bias due to the positioning of cylinders (see also \citealp{mukae17}).


\section{Results}



\begin{figure*}[h!]
	\begin{center}
	\plotone{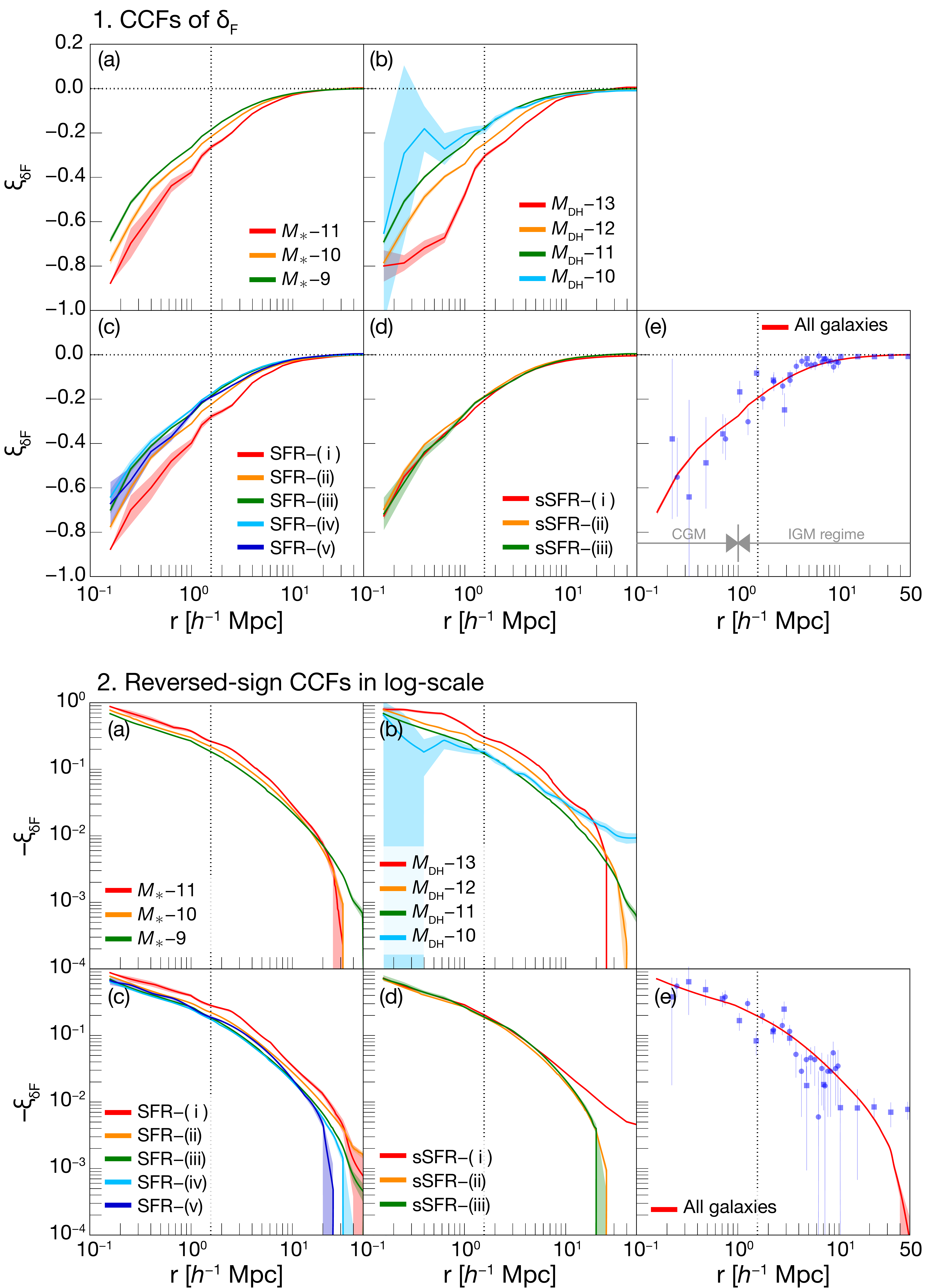}
	\caption{
    1): CCFs obtained from our simulation as a function of  radius in comoving units. 
    Dashed vertical and horizontal lines represent a half of the pixel size in transverse direction and a mean separation of LoS. The definition of the CGM and IGM regimes used in this paper is also shown in $Panel$
    {\it (e).}
    {\it Panels (a)--(d)}: CCFs for each subsample divided by $\Mstar$, $M_\text{DH}$, SFR, and sSFR. {\it Panel (e)}: CCF calculated using all galaxies. 
    Blue circles and squares indicate the observational estimates of CCF between $\delta_\text{F}$ and LBGs from \citet{adelberger05} and \citet{bielby17}, respectively. 2): 
    Sign-flipped CCF of Figure~\ref{fig:IM_all_CCF}--1
    is plotted in log-scale. The vertical dashed line is the same as in Figure~\ref{fig:IM_all_CCF}--1.
    }
	\label{fig:IM_all_CCF}
	\end{center}
\end{figure*}

\begin{figure*}
	\begin{center}
	\includegraphics[width=\linewidth]{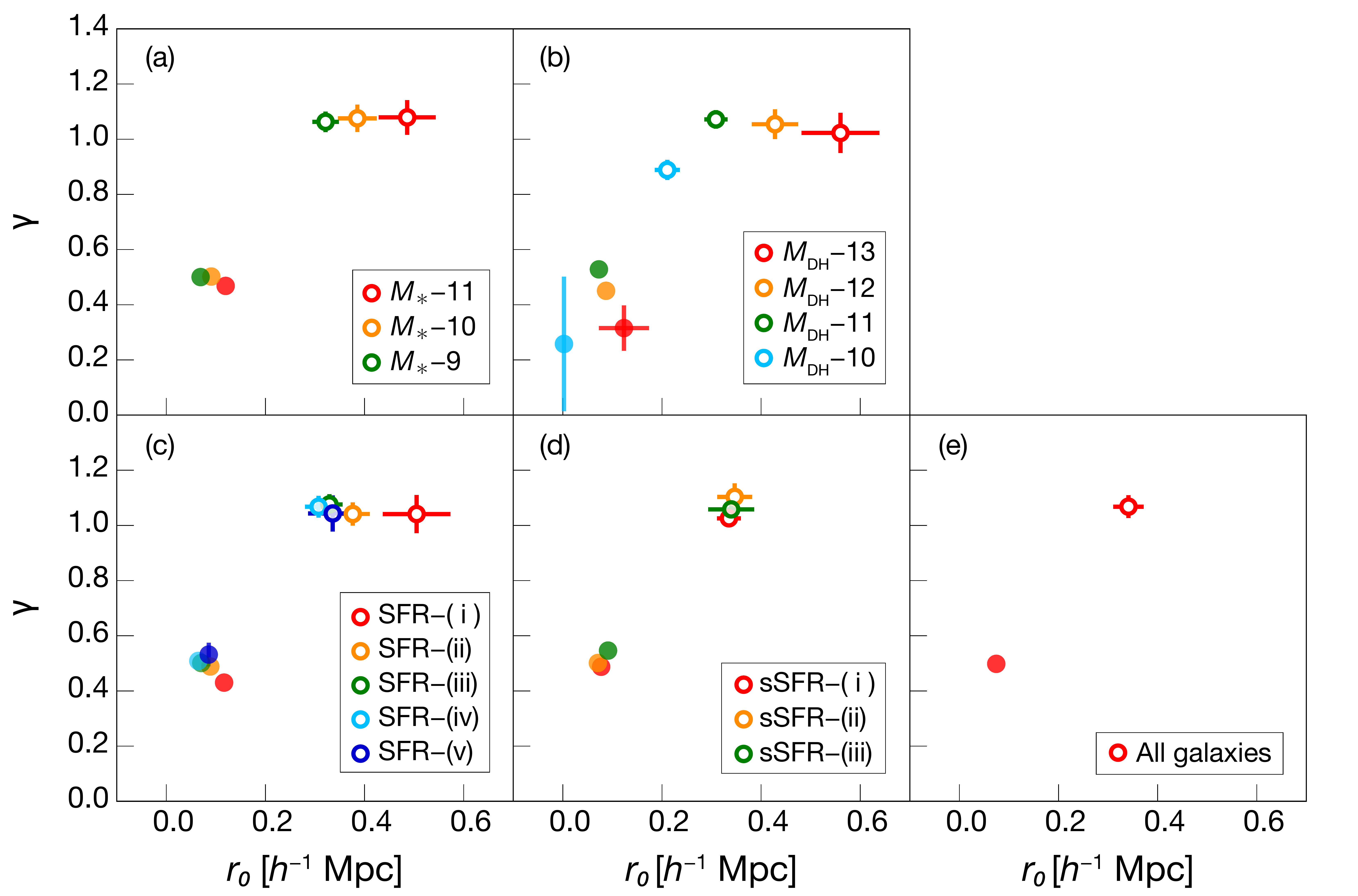}
	\caption{
    Best-fit parameters of a power-law fitting. Filled and open circles represent the best-fit parameters obtained from the CGM and IGM regime, respectively. When error bars are not recognized, they are smaller than symbol sizes. 
    Panels (a)--(e): the parameters for the CCFs in $\Mstar$, $M_\text{DH}$, SFR, and sSFR categories, and for the CCF from all galaxies. 
    }
	\label{fig:fit_IM_all_CCF}
	\end{center}
\end{figure*}


\subsection{Cross-correlation Analysis}

\subsubsection{Ly$\alpha$ Absorption Fluctuation}
\label{sec:result_ccf}

The CCFs of all galaxies and subsamples are shown in Figure \ref{fig:IM_all_CCF}--1. 
We also present the sign-flipped CCF plotted in log-scale  in Figure \ref{fig:IM_all_CCF}--2 for the discussion in Sections \ref{sec:result_ccf_gas} and \ref{sec:dis_CCF_origin}.
We detect a CCF signal up to $r\sim40$ $h^{-1}$ Mpc, which is in good agreement with observations by \citet{adelberger05} and \citet{bielby17}. 

Further investigations of CCF for four subsamples (divided by $\Mstar$, $M_\text{DH}$, SFR, and sSFR) are presented in Figure\,\ref{fig:IM_all_CCF}--1(a--d).
Most of the CCFs show monotonic increase from the center to $r=20-50$\,$h^{-1}$\,Mpc, except for the $M_\text{DH}$--$13$ and $M_\text{DH}$--$10$ sample which show irregular shapes.
Considering our tests for CCF reproducibility by a small sample size in Appendix \ref{app:number}, a swelling at $r\sim0.4$ $h^{-1}$ Mpc in $M_\text{DH}$--$10$ can be  attributed to the small sample size. 
While for $M_\text{DH}$--$13$, we regard a loosely bump at $r=0.3-0.8$ $h^{-1}$ Mpc as a real feature. 

We find that there is a clear tendency of CCF signal depending on the subsample, except for sSFR subsample. It is that the CCF signal becomes stronger with increasing galaxy masses and SFRs, but SFR--(v) subsamples do not follow this trend. 
A turnover radius where $\xi_\text{$\delta$F}$ reaches about zero also shows a trend for galaxies in $\Mstar$ and $M_\text{DH}$ subsamples (see Figure \ref{fig:IM_all_CCF}--2(a) and (b)), that a sample with a higher CCF signal drops rapidly to zero at a smaller radius, and hereafter we call this as `turnover radius'. 
For SFR samples, however, the CCF signal does not seem to correlate with turnover radius. 
On the other hand, sSFR samples do not show any obvious trend in their CCFs. Nonetheless, the turnover radius of sSFR samples increases with increasing sSFR.
Likewise, in previous studies, \citet{turner17} have demonstrated the halo mass dependency of the median $\tau_\text{{\sc Hi}}$ as a function of distance from their modeled galaxies.
\citet{meiksin17} have investigated  $\delta_\text{F}$ for galaxies in $M_\text{DH}$ subsamples against projected impact parameter and found an increase in $\delta_\text{F}$ with halo mass.
Observationally, \citet{chen05} have measured two-point cross-correlation $\xi_\text{ga}$ between Ly$\alpha$ absorbers and absorption-line-dominated or emission-line-dominated galaxies which are presumably massive early-type and star-forming galaxies, and presented a different amplitude of $\xi_\text{ga}$ (see also \citealp{chen09}). 
\citet{wiliam07} have also possibly found differences in the cross-correlation signal between absorption-line-dominated and emission-line-dominated galaxies (see their Fig. 4), although they have not claimed that it is a statistically significant detection.

To characterize our CCFs, we fit them with a power-law of
\begin{equation}
    \xi_\text{$\delta$F}(r) = \left( \frac{r}{r_0} \right)^{-\gamma},
\end{equation} 
where $r_0$ and $\gamma$ are a clustering length and slope. 
We apply the power-law fitting to the CCFs in Figure~\ref{fig:IM_all_CCF}--2
over $0.1 - 1$ $h^{-1}$ Mpc and
$1-10$ $h^{-1}$ Mpc separately, corresponding to the CGM and IGM regimes. 
Note that $r_0$ and $\gamma$ reflect the amplitude and slope of a CCF, respectively.
Best-fit parameters of the all CCFs are presented in Figure \ref{fig:fit_IM_all_CCF}. Filled and open circles represent the best-fit parameters of the CGM and IGM regimes, respectively. 

For all galaxies, we obtain the best-fit parameters of 
($r_0$, $\gamma$) $=$ ($0.07\pm0.004$, $0.50\pm0.01$) and 
($0.34\pm0.03$, $1.07\pm0.04$) for the CGM and IGM regimes (see also Figure \ref{fig:fit_IM_all_CCF}(e)).
Several observational studies have performed a power-law fitting to their CCFs between Ly$\alpha$ absorption and galaxies.
\citet{tumm14} have calculated a CCF between Ly$\alpha$ absorption and LBGs at $z\sim3$, and fit it by a double power-law, showing 
($r_0$, $\gamma$) $=$ ($0.08\pm0.04$, $0.47\pm0.10$) and 
($0.49\pm0.32$, $1.47\pm0.91$) for the CGM and IGM regimes used at $r=1.6$ $h^{-1}$ Mpc as a border.
A subsequent study of the IGM--LBG clustering by \citet{bielby17} have been described their CCF by a single power-law with 
($r_0$, $\gamma$) $=$ ($0.27\pm0.14$, $1.1\pm0.2$). 
A power-law fitting to a CCF of weak H\,{\sc i} ($N_\text{H\,{\sc i}}<10^{14}$ cm$^{-2}$) IGM and galaxies at $z<1$ have been also attempted, and resulted in 
($r_0$, $\gamma$) $=$ ($0.2\pm0.4$, $1.1\pm0.3$) \citep{tejos14}. 
Although the fitting range for power-law fitting is different among studies, our best-fit parameters for both CGM and IGM regimes are comparable to those previous observations within the error.

We next show best-fit parameters obtained from all categories. 
For $\Mstar$ and $M_\text{DH}$, they show similar trend in both $r_0$ and $\gamma$ of the both CGM and IGM regimes. A clustering length $r_0$ becomes longer with increasing mass independent of the regimes. 
We remind the reader that $r_0$ reflects the amplitude of the CCFs. Thus, we can confirm particularly in Figure~\ref{fig:IM_all_CCF}--2 that the higher the mass becomes, the stronger the CCF amplitude is.
Meanwhile, the slope $\gamma$ becomes smaller with increasing mass in the CGM regime, but is consistent with being constant within the error in the IGM regime.
In the literature, \citet{tumm14} have calculated the mass-dependent CCFs by dividing their simulated galaxies into two categories of $\Mstar>10^8$ and $\Mstar>10^9$ $h^{-1}$\ M$_\odot$. Their double power-law fit for these $\Mstar>10^8$ and $\Mstar>10^9$\,$h^{-1}$\,M$_\odot$ samples have been given 
($r_0$, $\gamma$) $=$ ($0.10\pm0.07$, $0.46\pm0.22$) and ($0.16\pm0.09$, $0.46\pm0.19$) for the CGM regime, and
($0.51\pm0.39$, $1.25\pm0.61$) and ($0.61\pm0.34$, $1.18\pm0.43$) in the IGM regime.
Although differences of $r_0$ and $\gamma$ between $\Mstar>10^8$ and $\Mstar>10^9$\,$h^{-1}$\,M$_\odot$ samples are within the error, the similar trend is confirmed in $r_0$.
For the SFR sample except SFR--(v), we identify that $r_0$ becomes greater with increasing SFR. It is naturally explained by the fact that our galaxy sample are mostly on the star formation main sequence between SFR and stellar mass. On the other hand, $\gamma$ has no clear trend in either the CGM or IGM regime. 
Although the CCFs of all three sSFR category are comparable as presented in Figure \ref{fig:IM_all_CCF}--1(d), best-fit parameters of both $r_0$ and $\gamma$ are different each other. Interestingly, both $r_0$ and $\gamma$ show identical trends in the both CGM and IGM regimes, that sSFR--(i) shows the smallest value. 
We briefly discuss its reason in Sections \ref{sec:dis_CCF_origin} and \ref{sec:dis_oe}.



\begin{figure*}
	\begin{center}
	\plotone{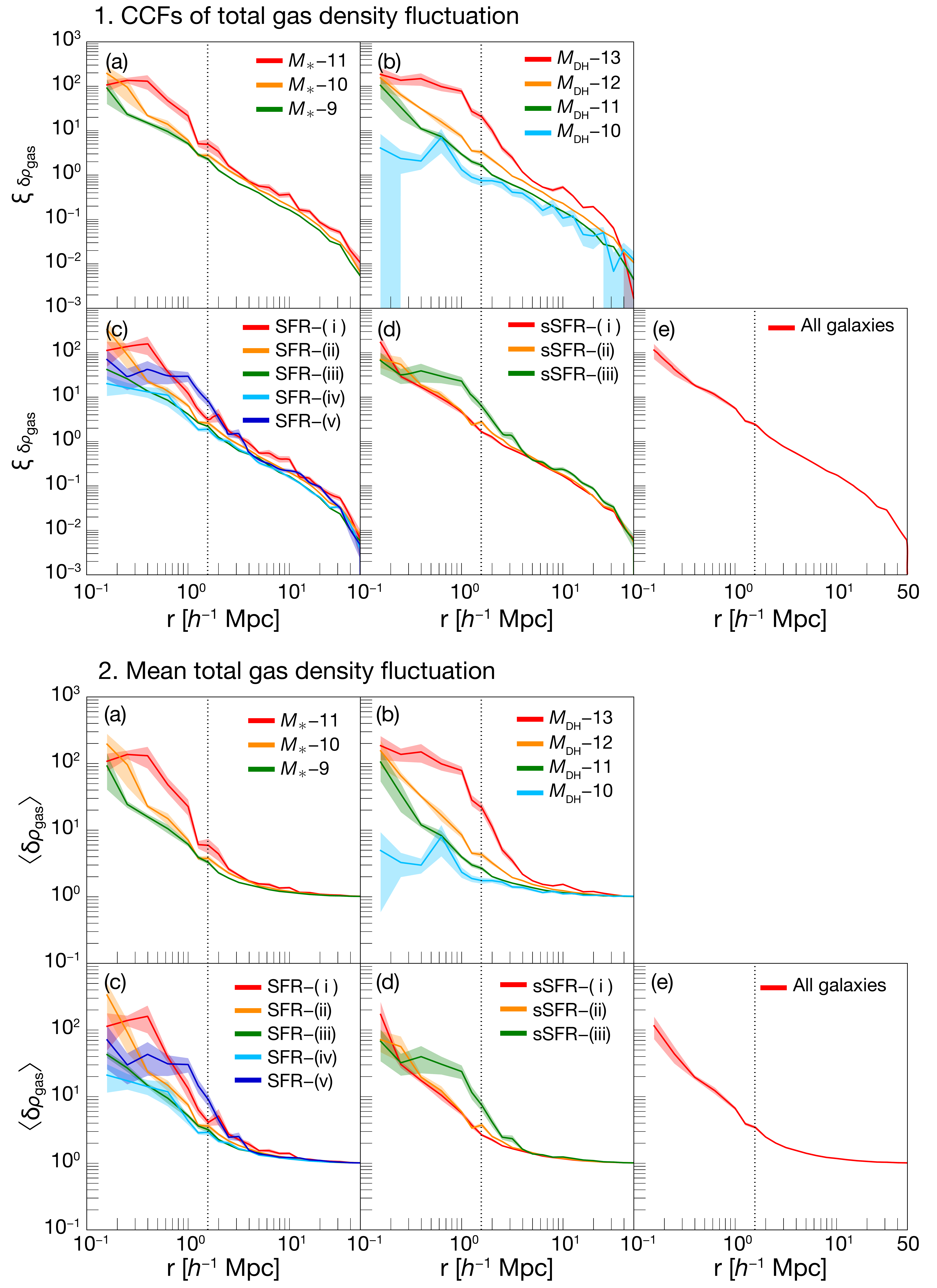} 
	\caption{
    1) CCFs of total gas density fluctuation ($\delta_{\rho_\text{gas}}=\rho_\text{gas}$/$\langle \rho_z \rangle$) as a function of comoving distance from galaxies. 2) Mean total gas density fluctuation around galaxies.
    }
	\label{fig:IM_tgas}
	\end{center}
\end{figure*}

\begin{figure*}
	\begin{center}
	\plotone{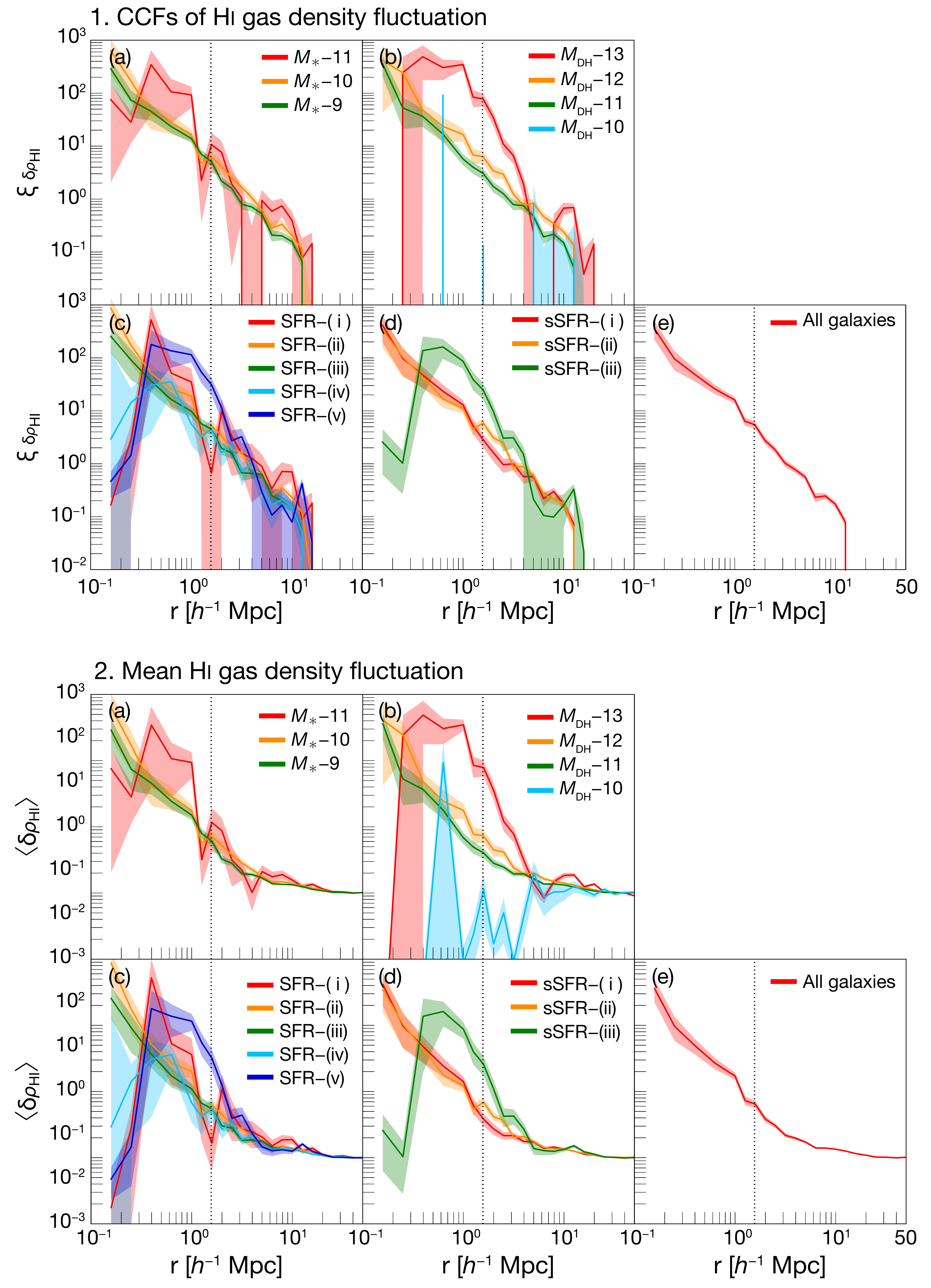} 
	\caption{
    1) CCFs of {\sc Hi} density fluctuation ($\delta_{\rho_\text{{\sc Hi}}}=\rho_\text{{\sc Hi}}$/$\langle \rho_z \rangle$) as a function of comoving distance from galaxies. 2) Mean {\sc Hi} density fluctuation around galaxies.
    }
	\label{fig:IM_hgas}
	\end{center}
\end{figure*}


\subsubsection{Gas Density Fluctuation}
\label{sec:result_ccf_gas}

A $\delta_\text{F}$ value of LoS data shall correlate with {\sc Hi} gas density at the position. Thus, a variety of the CCFs amplitudes can be attributed to a variety of the local gas density around galaxies. 
To verify the hypothesis, we evaluate CCFs of gas density fluctuations around galaxies defined by 
\begin{equation}
    \delta_{\rho} \equiv \frac{\rho}{\langle \rho_z \rangle},
\end{equation}
\begin{equation}
    \xi_\text{$\delta\rho$} = \frac{1}{N(r)} \sum_{i=1}^{N(r)} \delta_{\rho_{g,i}} - \frac{1}{M(r)} \sum_{j=1}^{M(r)} \delta_{\rho_{ran,j}},
\end{equation}
where $\delta_{\rho}$ is the gas density fluctuation defined by the ratio of a gas density at one LoS pixel $\rho$ to the mean gas density at each redshift of $\Delta z=0.01$, $\langle \rho_z \rangle$; $\xi_\text{$\delta\rho$}$ is the CCF; $\delta_{\rho_{g},i}$ and $\delta_{\rho_{ran},j}$ are the gas density fluctuation for the pixel $i$ and $j$ at the distance $r$ from galaxies and random points.
Similarly, in Equation (5), $N(r)$ and $M(r)$ are the numbers of pixel-galaxy and pixel-random pairs in the bin with the distance $r$.
The CCF of gas density fluctuations are measured for both total gas and {\sc Hi}  ($\delta_{\rho_\text{gas}}$ and $\delta_{\rho_\text{{\sc Hi}}}$), and are shown in Figures \ref{fig:IM_tgas}--1 and \ref{fig:IM_hgas}--1.
For the comparison to the CCFs of $\delta_\text{F}$, the log-scale inversion CCFs of $\delta_\text{F}$ (i.e., $-\log \delta_\text{F}$) are also shown in Figure \ref{fig:IM_all_CCF}--2. 
Due to several negative values in $\delta_{\rho_\text{{\sc Hi}}}$, we also present the mean gas density fluctuations around galaxies in Figures \ref{fig:IM_tgas}--2 and \ref{fig:IM_hgas}--2. 

First, we start from the CCFs of total gas density fluctuation $\delta_{\rho_\text{gas}}$, in Figure \ref{fig:IM_tgas}. 
Overall trends for each category (i.e., $\Mstar$, $M_\text{DH}$, SFR, and sSFR) are almost the same as that of $\delta_\text{F}$'s CCFs.
It is that overall CCFs are monotonically decreasing with radius, and 
a CCF signal becomes higher with increasing galaxies mass (either $\Mstar$ or $M_\text{DH}$) or SFR. 
However several samples show a notable behavior in their CCFs in the context of a CCF's strength or shape at a certain radius. 
For $\Mstar$--$11$, $M_\text{DH}$--$13$ and SFR--(i), we find a slight decline of $\xi_{\delta\rho_\text{gas}}$ at the center. It indicates the decline of relative total gas density in the proximity of those galaxies.
For SFR--(v) and sSFR--(iii), their CCFs show a convex feature at $r=0.3-2$ $h^{-1}$ Mpc and even have a highest signal among each category over that radius.
It is note for $M_\text{DH}$--10 that a swelling at $r\sim0.5$ $h^{-1}$ Mpc can be due to its small sample size. 

The CCFs of {\sc Hi} gas density fluctuation $\delta_{\rho_\text{{\sc Hi}}}$ are shown in Figure \ref{fig:IM_hgas}. 
Compared to the {\sc Hi} optical depth distribution on the LoS (Figure \ref{fig:IM_all_CCF}--1, $\delta_\text{F}$), raw {\sc Hi} gas particle has slightly discrete distribution (Figure \ref{fig:IM_hgas}--2, $\delta_{\rho_\text{{\sc Hi}}}$). 
This is because that we consider the line broadening based on the Voigt profile (see also Equation 2). 
As a result, the CCFs in $\delta_\text{F}$ have more smooth shape than that in $\delta_{\rho_\text{{\sc Hi}}}$. 
Likewise to overall trends seen in the CCFs of $\delta_{\rho_\text{gas}}$, we find that a CCF signal becomes higher as increasing $\Mstar$, $M_\text{DH}$ or SFR in general. The trend is also about the same as one of found in CCFs of $\delta_\text{F}$. The consistency of CCFs' trends is naturally explained by considering the Equations (2) and (3) that $\delta_\text{F}$ proportionals to {\sc Hi} number density.
We also find that a similar irregular CCF identified in CCFs of $\delta_{\rho_\text{gas}}$ is seen in several samples: a decline of $\xi_{\delta\rho_\text{{\sc Hi}}}$ at the center in $\Mstar$--$11$, $M_\text{DH}$--$13$ and SFR--(i), and a convex profile and strongest signal at $r=0.3-2$ $h^{-1}$ Mpc in SFR--(v) and sSFR--(iii). In addition to above irregular CCFs' shapes, SFR--(v) and sSFR--(iii) show a significant decline of $\xi_{\delta\rho_\text{{\sc Hi}}}$ value at the center. It suggests that {\sc Hi} densities around galaxies in $\Mstar$--$11$, $M_\text{DH}$--$13$, SFR--(i), SFR--(v) or sSFR--(iii) are also relatively low in general. 
We discuss it in Section \ref{sec:dis_oe}.


\begin{figure*}
	\begin{center}
	\includegraphics[width=\linewidth]{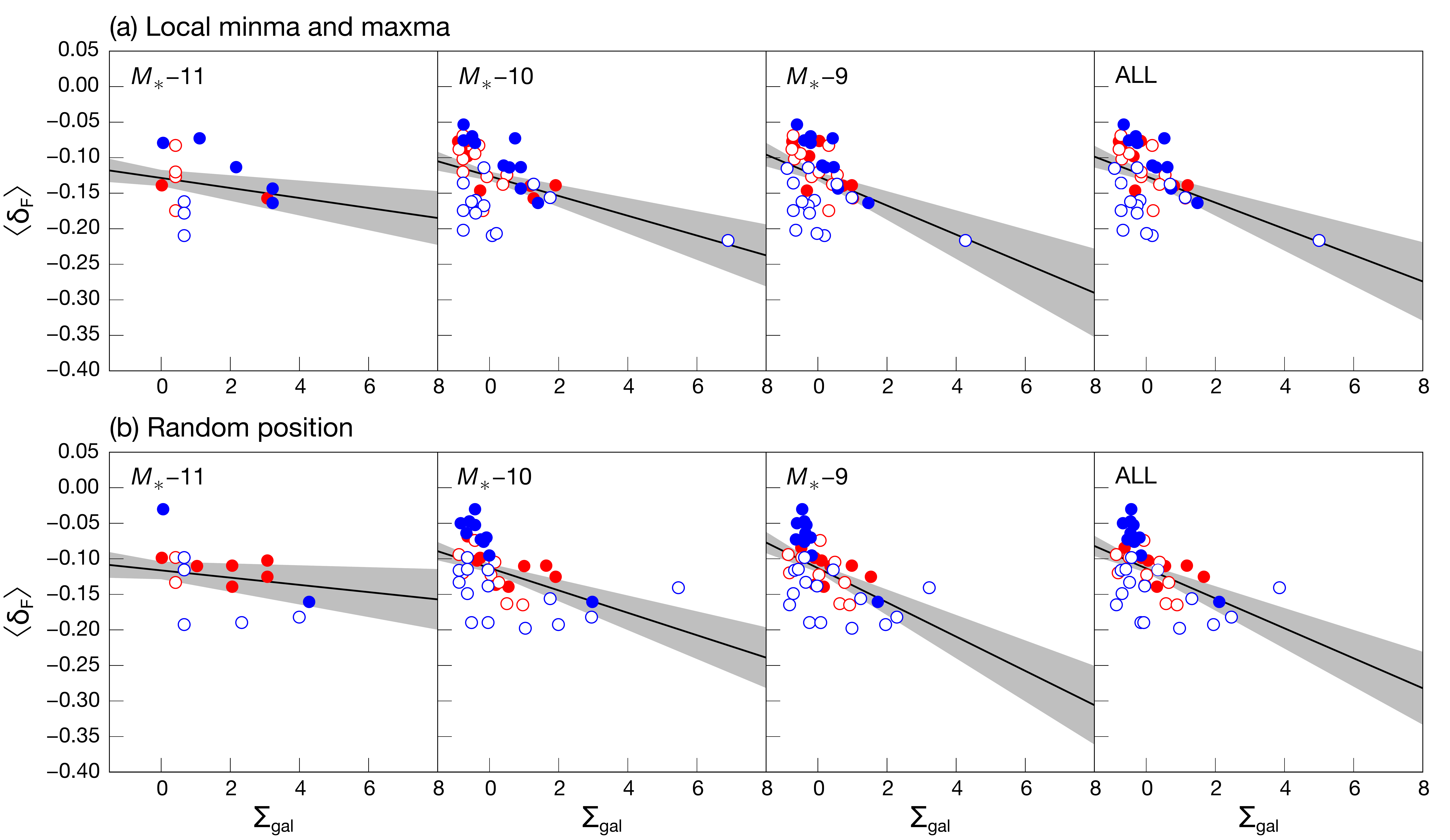} 
	\caption{
    Overdensity analysis obtained from (a) local minima and maxima of gas density field, and (b) random points, respectively. Results for $\Mstar$--$11$, $\Mstar$--$10$, $\Mstar$--$9$ and ALL are shown from left to right. Open and filled circles colored in red and blue indicate data points from each of four 2D LoS map. 
    The best-fit linear regression is shown in black line with its errors shown by grey shade. 
    }
	\label{fig:oe}
	\end{center}
\end{figure*}



\begin{deluxetable*}{l|rrrr|rr|rr}
\tablenum{2}
    \tablecaption{Measurement results of over-density analysis}
\startdata
\tablehead{
    \colhead{Sample Name}
    & \colhead{$N^{(1a)}$} & \colhead{$N^{(1b)}$} & \colhead{$N^{(1c)}$} & \colhead{$N^{(1d)}$}
    & \colhead{$R_s^{(2)}$} & \colhead{$p^{(2)}$}
    & \colhead{$\alpha^{(3)}$} & \colhead{$\beta^{(3)}$}
}
    $\Mstar$--$11$ 
    & 134 & 139 & 100 & 85
    & $-0.41$ & $0.12$ & $-0.129\pm0.011$ & $-0.007\pm0.003$    \\
    $\Mstar$--$10$
    & 1709 & 1555 & 1226 & 1182
    & $-0.42$ & $6.27$e$-3$ & $-0.126\pm0.006$ & $-0.014\pm0.005$   \\
    $\Mstar$--$9$  
    & 4658 & 4356 & 3984 & 3921
    & $-0.33$ & $0.03$ & $-0.126\pm0.006$ & $-0.020\pm0.007$    \\
    ALL & 6501 & 6050 & 5310 & 5188
    & $-0.37$ & $0.02$ & $-0.126\pm0.006$ & $-0.018\pm0.006$    \\
\hline\hline
    & -- & -- & -- & -- & 
    $R_s^{(4)}$ & $p^{(4)}$ & $\alpha^{(5)}$ & $\beta^{(5)}$ \\
\hline
    $\Mstar$--$11$
    & 134 & 139 & 100 & 85
    & $-0.54$ & $0.02$ & $-0.116\pm0.012$ & $-0.005\pm0.004$ \\
    $\Mstar$--$10$
    & 1709 & 1555 & 1226 & 1182
    & $-0.57$ & $8.46$e$-5$ & $-0.113\pm0.006$ & $-0.016\pm0.005$ \\
    $\Mstar$--$9$
    & 4658 & 4356 & 3984 & 3921
    & $-0.53$ & $2.48$e$-4$ & $-0.113\pm0.006$ & $-0.024\pm0.006$ \\
    ALL & 6501 & 6050 & 5310 & 5188
    & $-0.51$ & $3.70$e$-4$ & $-0.114\pm0.006$ & $-0.021\pm0.006$ \\
\enddata
\tablecomments{
		$^{(1)}$ Number of galaxies in (1a) $2.07<z<2.102$, (1b) $2.215<z<2.247$, (1c) $2.3<z<2.332$ and (1d) $2.45<z<2.482$.
		$^{(2)}$ Spearman's coefficient and $p-$value. The $\langle \delta_\text{F} \rangle - \Sigma_\text{gal}$ relation is examined around local minima and maxima. 
		$^{(3)}$ The best-fit parameters of chi-square fitting of the $\langle \delta_\text{F} \rangle - \Sigma_\text{gal}$ relation examined around local minima and maxima. 
		$^{(4)}$ Spearman's coefficient and $p-$value. The $\langle \delta_\text{F} \rangle - \Sigma_\text{gal}$ relation is examined around random points. 
		$^{(5)}$ The best-fit parameters of chi-square fitting of the $\langle \delta_\text{F} \rangle - \Sigma_\text{gal}$ relation examined around random points. 		
}
\end{deluxetable*}
\label{tab:overdensity}


\subsection{Overdensity Analysis}
\label{sec:res_oe}

We present the results of overdensity analysis in Figures~\ref{fig:oe}--(a) and \ref{fig:oe}--(b), which are derived from local minima/maxima and random positions in the gas density field.    
The analysis is performed for all galaxies and $\Mstar$-dependent subsamples of $\Mstar$--$11$, $\Mstar$--$10$, and $\Mstar$--$9$. 
We evaluate $\Sigma_\text{gal}$ and $\langle \delta_\text{F} \rangle$ values in each of four redshift slices indicated by the open or filled circles colored in red or blue. Exact galaxy counts used in each redshift slice is shown in Table~\ref{tab:overdensity}. 

First, we find possible anti-correlations between $\Sigma_\text{gal}$ and $\langle \delta_\text{F} \rangle$ in Figure \ref{fig:oe}--(a). To statistically assess those correlations, we perform Spearman's rank correlation test, and  obtain correlation coefficients ranging from $R_s=-0.33$ to $R_s=-0.42$, indicating a mild anti-correlation.
Similarly, \citet{mukae17} also identified a mild anti-correlation in their $\langle \delta_\text{F} \rangle$--$\Sigma_\text{gal}$ distribution with $R_s=-0.39$. 

We should remark about the effect by the outlier data points in $\Mstar$--10, $\Mstar$--9 and ALL of Figure~\ref{fig:oe}--(a). 
We repeat the Spearman's rank correlation test for all data points but without the outlier, and obtain $R_s = (-0.37, -0.28, -0.32)$ for ($\Mstar$--10, $\Mstar$--9, ALL) samples. Therefore, weak anti-correlations are still confirmed even without the outliers.

To characterize the $\langle \delta_\text{F} \rangle$--$\Sigma_\text{gal}$ distribution, we follow \citet{mukae17} and apply chi-square fitting in Figure \ref{fig:oe} with the linear model of 
\begin{equation}
  \langle \delta_\text{F} \rangle = \alpha + \beta\, \Sigma_\text{gal}.
\end{equation}
The best-fit parameters of $\alpha$ and $\beta$ are summarized in Table \ref{tab:overdensity}. 
We find that $\alpha\sim-0.13$, which is about the same for all of four samples within the error, while $\beta$ becomes slightly larger with increasing $\Mstar$, although they are still similar within the error: $\beta=-0.007\pm0.003$, $-0.014\pm0.005$, $-0.020\pm0.007$ for $\Mstar$--11, $\Mstar$--10, $\Mstar$--9, respectively. 

Note that the best-fit parameters of anti-correlations for $\Mstar$--10, $\Mstar$--9 and ALL without outliers appear to be comparable within the error.
We compare the best-fit parameters of the `ALL' sample to those in \citet{mukae17}, which are ($\alpha$, $\beta$) $=$ ($-0.17\pm0.06$, $-0.14_{-0.16}^{+0.06}$). 
We find a similarly in $\alpha$ but a larger difference in $\beta$, showing a much shallower slope for our sample. 
The shallower slope of simulated galaxy sample has also been found in \citet{Nagamine20}.
It may be attributed to photo-$z$ errors in the observational data. 
If photo-$z$ errors are large, then some galaxies would contaminate the sample, and the value of $\Sigma_\text{gal}$ would be smeared out. 
As a result, the observed $\Sigma_\text{gal}$  only has a narrow dynamic range, which could make the apparent correlation steeper than the real one.

We also examine the result of overdensity analysis based on randomly-selected points in order to verify the effect of position bias (Figure~\ref{fig:oe}--(b)). 
The Spearman's rank correlation tests for randomly-selected positions yield mild anti-correlations with $R_s=-0.51$ to $R_s=-0.57$.
The best-fit parameters of the linear model ($\alpha$ and $\beta$) are also comparable to those from Figure~\ref{fig:oe}--(a) within the error. 
Moreover, the trend found in best-fit $\alpha$ and $\beta$ as a function of stellar-mass is also confirmed. 
Therefore, we conclude that the position bias of overdensity analysis is not affecting the $\langle \delta_\text{F} \rangle$--$\Sigma_\text{gal}$ correlation seriously.


\section{Discussion}


\subsection{Origin of CCF Variations}
\label{sec:dis_CCF_origin}

We presented in Section~\ref{sec:result_ccf} that the CCF of $\delta_\text{F}$ varies depending on galaxy mass and SFR. 
To find the origin of its variation, we also calculated the CCFs of $\delta_{\rho_\text{gas}}$ and $\delta_{\rho_\text{{\sc Hi}}}$ in Section \ref{sec:result_ccf_gas}, and found that their signal strengths also depend on $\Mstar$, $M_\text{DH}$ and SFR. 
It suggests that different relative gas density around galaxies is causing the variation of the CCFs of $\delta_\text{F}$. 
Considering the relation between $\delta_\text{F}$ and {\sc Hi} number density in Equations (2)--(4), a similar trend of CCFs in $\delta_\text{F}$ and $\delta_{\rho_\text{{\sc Hi}}}$ is reasonable. 
The same trend even for $\delta_{\rho_\text{gas}}$ probably means that the total gas density correlates with {\sc Hi} gas density in general. 
Therefore, we argue that the variation of  $\delta_\text{F}$ CCF is caused by different gas distribution around galaxies. 

We find that not only the $\delta_\text{F}$ CCFs, but also their best-fit parameters of power-law fitting vary depending on galaxies mass and SFR (see also Figure~\ref{fig:fit_IM_all_CCF}). 
Particularly, our best-fit clustering lengths $r_0$ for the IGM regime show an increase with increasing mass or SFR of galaxies.
A dependency of a CCF signal strength and its best-fit parameters on {\sc Hi} column density ($N_\text{{\sc Hi}}$) of CGM and IGM has also been reported in observational studies. 
For example, \citet{ryan-weber06} has calculated CCFs by splitting their absorber sample into two based on absorber's $N_\text{{\sc Hi}}$, and found that high--$N_\text{{\sc Hi}}$ subsample shows stronger correlation than low--$N_\text{{\sc Hi}}$ subsample.
\citet{tejos14} have calculated CCFs depending on $N_\text{{\sc Hi}}$ for galaxies at $z<1$, and found a positive correlation between best-fit parameters (both $r_0$ and $\gamma$) and $N_\text{{\sc Hi}}$. They have also demonstrated a dramatic change of CCF signals depending on $N_\text{{\sc Hi}}$, showing more than a factor of ten higher CCF signal in $N_\text{{\sc Hi}}\geq10^{14}$ cm$^{-2}$ sample compared to that of $N_\text{{\sc Hi}}<10^{14}$ cm$^{-2}$ sample.
Similarly, \citet{bielby17} have analyzed cross-correlation between Ly$\alpha$ absorption with different $N_\text{{\sc Hi}}$ measurements and LBGs at $z=3$, and presented a positive correlation between best-fit parameters of their CCFs and $N_\text{{\sc Hi}}$. 
These studies have also argued for the relation between the best-fit parameters ($r_0$ and $\gamma$) and gas, particularly {\sc Hi}. 
Larger $r_0$ imply stronger clustering of {\sc Hi} systems around galaxies.
On the other hand, for $\gamma$, previous studies have suggested the necessity for additional baryonic physics to explain its changes with $N_\text{{\sc Hi}}$. 

Considering the above discussion, the signal strength and best-fit parameters ($r_0$ and $\gamma$) of CCF depends on relative gas densities on Mpc-scale near the galaxy.  
If the gas has a high density and clusters around a galaxy, the resultant CCF between Ly$\alpha$ absorbers and the galaxy must have a higher signal and give larger best-fit parameters for the IGM-regime.
The variation of CCFs is hence attributed to different gas density around each galaxy and the strength of galaxy--IGM connection.


\subsection{Which Type of Galaxies Strongly Correlate with the IGM?}
\label{sec:dis_oe}

Under the $\Lambda$CDM paradigm, massive galaxies are expected to strongly correlate with the underlying dark matter (e.g., \citealp{mo02,zehavi05}), and hence with IGM as well.  
In that sense, more massive galaxies should strongly cluster in higher density regions compared to less massive galaxies. 
In Section \ref{sec:result_ccf}, we confirmed that the CCF signal becomes stronger with increasing mass (both $\Mstar$ and $M_\text{DH}$) of a galaxy. In addition, we also find that the turnover radius of CCFs becomes smaller with mass in $\Mstar$--11 and $M_\text{DH}$--13, which is likely to be the result of stronger connection between massive galaxies and higher density regions. 

From the overdensity analysis in Section \ref{sec:res_oe}, we find that the slope of the anti-correlation between $\langle \delta_\text{F} \rangle$ and $\Sigma_\text{gal}$ becomes shallower with increasing M$_\star$, although its difference is still within the error. 
A shallower slope in $\Mstar$--11 sample indicates  stronger clustering of massive galaxies around dense {\sc Hi} IGM. 

Above results from both methods (CCF and overdensity analysis) imply that massive galaxies are strongly clustered in high-density regions in the cosmic web, while less massive galaxies have an opposite trend.

\begin{figure}
	\begin{center}
	\plotone{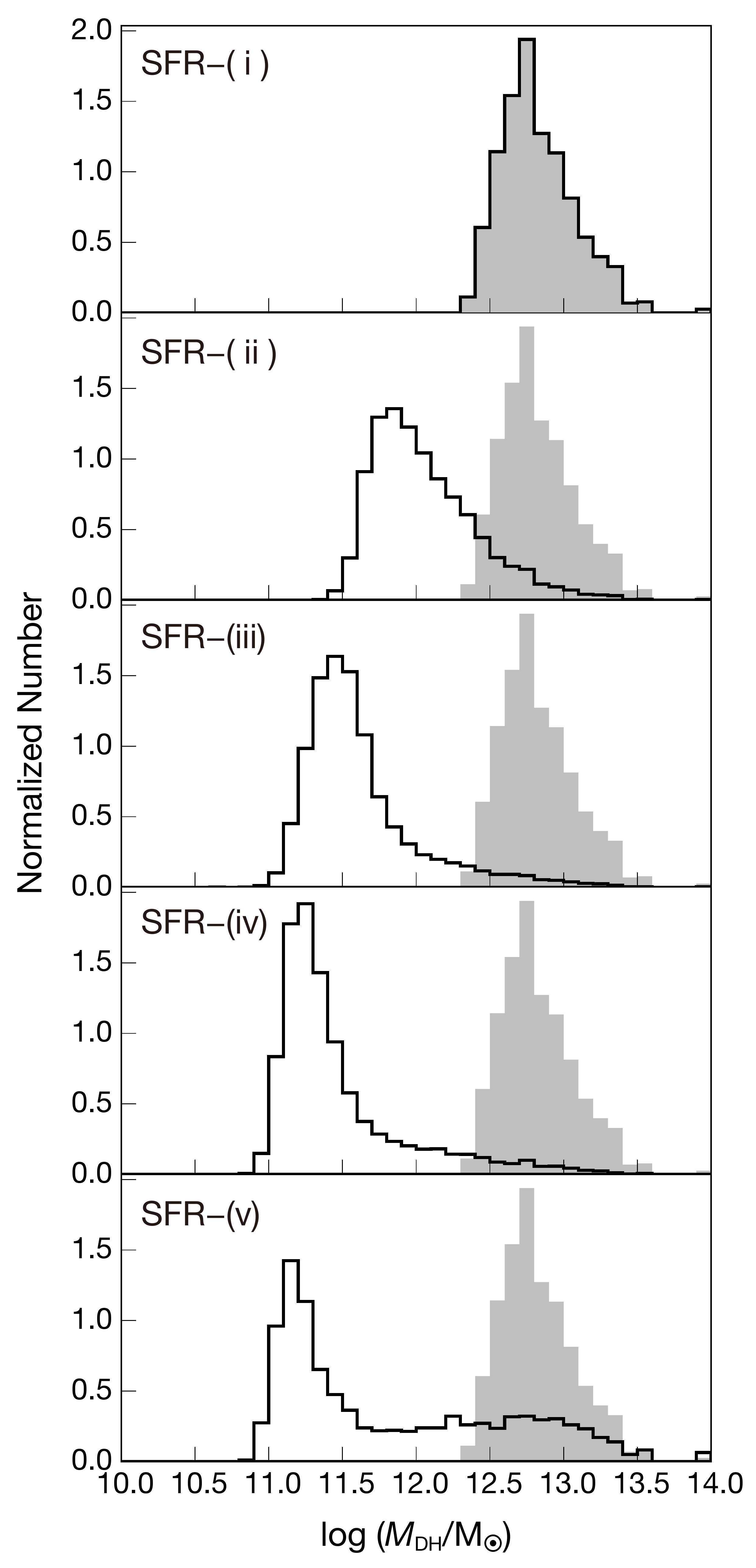} 
	\caption{
    Normalized number histogram of galaxies as a function of $M_{\rm DH}$ for each SFR subsample. 
    The solid black line indicates the histogram of each subsample, and the {SFR--(i)} histogram is overlaid in the bottom four panels as the gray-shaded histogram for comparison. 
    }
	\label{fig:hist_SFR_MDH}
	\end{center}
\end{figure}

The same trend should be true for SFR subsamples by considering the star-formation main sequence (e.g., \citealp{Brinchmann04,noeske07,elbaz07,daddi07,Speagle14,Schreiber15,Tomczak16}). The overall trend for the SFR samples is that the galaxies with higher (lower) SFRs correlate with higher (lower) gas density of the IGM. 
However, SFR--(v) subsample does not follow the trend and even seems to reside in the highest density among all SFR samples at $r=0.3-2$\,$h^{-1}$\,Mpc (see also Section \ref{sec:result_ccf_gas}). Such CCF behavior can be attributed to the halo mass distribution of galaxies in SFR--(v). Figure~\ref{fig:hist_SFR_MDH} represents the halo mass distribution of galaxies in all SFR samples. They generally have a single peak, but a mild bimodal distribution in SFR--(v), which has two peaks at $M_\text{DH}\sim10^{11}$ and $10^{12.5}-10^{13}$ M$_\odot$. It implies that the host dark matter halos of  SFR--(v) subsample can be roughly divided into two: one is less massive and the other is massive. 
Given the mass-dependency of IGM--galaxy connection, the strong CCF (i.e. high-density gas) seen at $r=0.3-2$\,$h^{-1}$\,Mpc of SFR--(v) subsample reflects high gas density around massive halos.

\begin{figure}
	\begin{center}
	\plotone{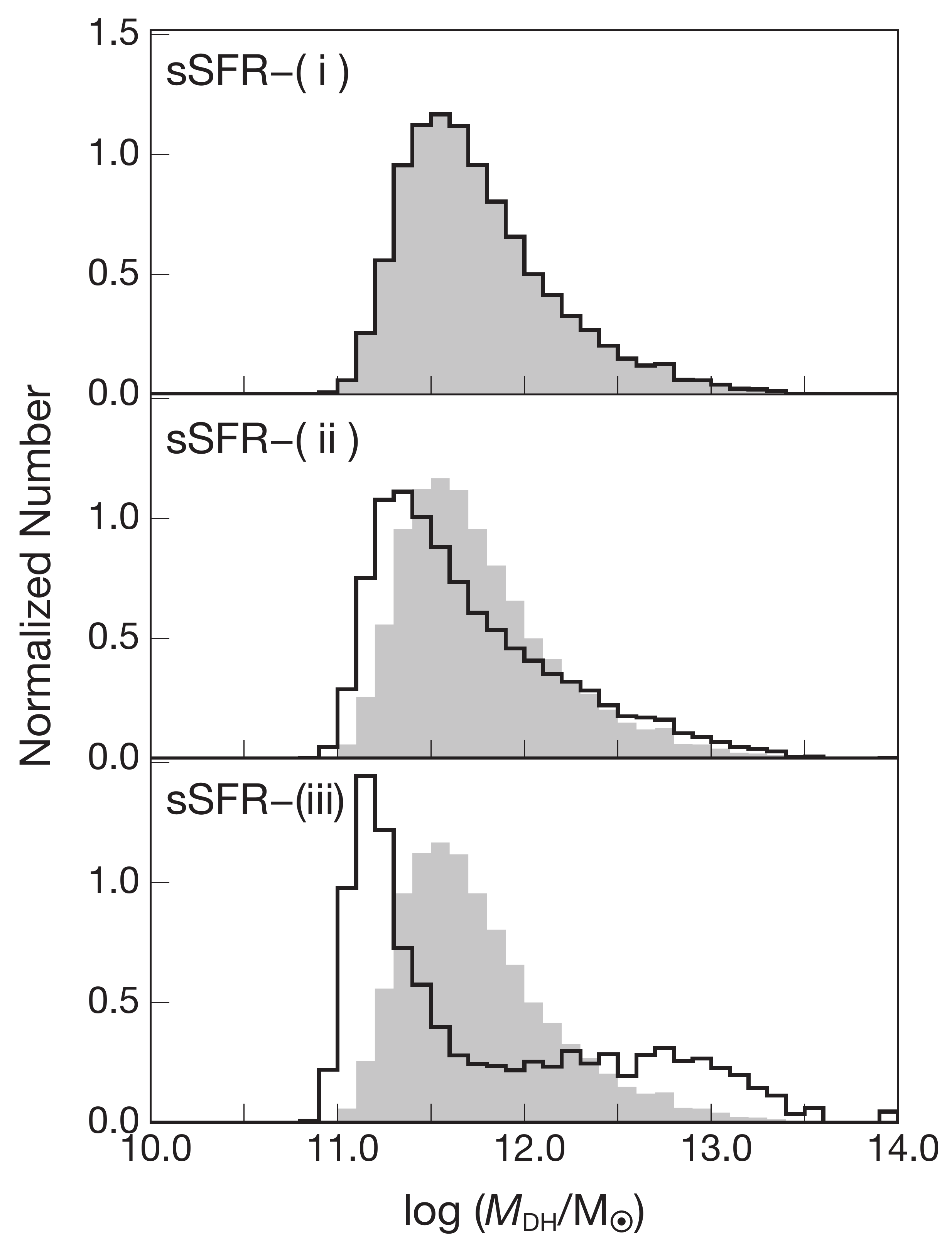} 
	\caption{
    Same as Figure~\ref{fig:hist_SFR_MDH}, but for sSFR subsamples. 
    The solid black line indicates the histogram of each subsample, and the sSFR--(i) histogram is overlaid in the bottom two panels as the gray-shaded histogram for comparison. 
    }
	\label{fig:hist_sSFR_MDH}
	\end{center}
\end{figure}

Our sSFR samples do not show any obvious trends in the CCF strength. We find that all three samples have comparable CCFs, although sSFR--(i) gives the smaller $r_0$ and $\gamma$ in both the CGM and IGM regimes.
It can be due to the same reason as what we see in the SFR samples. Similar to the SFR--(v) sample, the sSFR--(iii) sample has a mild bimodal halo mass distribution (see Figure~\ref{fig:hist_sSFR_MDH}). The halo mass distribution of sSFR--(ii) may have a tail extending to higher mass. As a result, these two samples possibly reside in higher density regions in the IGM than sSFR--(i). 

Summarizing the above discussions, we conclude that the dark matter halo mass is the most sensitive parameter to determine the baryonic environment around galaxies. 
Galaxies that are hosted by massive halos are generally located in high-density gaseous environment, resulting in a stronger signal of CCF.

Finally, we should briefly discuss about declining CCF signal at $r<0.3-0.4\,h^{-1}$\,Mpc around galaxies in $\Mstar$--$11$, $M_\text{DH}$--13, SFR--(i), SFR--(v), sSFR--(iii) subsamples (see Figs.\,\ref{fig:IM_tgas} and \ref{fig:IM_hgas}), which are hosted by the most massive halos among each category.
There are several possible reasons for the lack of {\sc Hi} gas in the central region of massive galaxies. 
The first possibility is that the gas particles in the central region are blown out to the CGM/IGM by SN feedback. 
The second possibility is that our feedback prescription without AGN contribution is still inadequate in pushing the gas away into the CGM/IGM for the massive galaxies due to their deep gravitational potential. 
As a result, most gas particles in the central region are consumed by star formation. 
We need more detailed analysis of our simulation to confirm the reason, but this is beyond the scope of this paper. 
We will try to address this issue in our future work. 
%



\subsection{Photo-$z$ vs. Spec-$z$ Data and the IGM--Galaxy Connection} 
\label{sec:dis_which}

In this subsection, we briefly discuss the reliability of using photo-$z$ and spec-$z$ data to investigate the IGM--galaxy connection. 
We also argue that some precautions must be taken when one applies the two analyses to observational data. 
We have demonstrated that the cross-correlation method succeeds in identifying the variations of CCF according to galaxy properties. 
The difficulty for the CCF method is the necessity of relatively accurate redshift measurements for galaxies. 
As we demonstrate in Appendix B, the CCF signal would be attenuated if the redshift uncertainty is large. 
According to the general photo-$z$ uncertainties at $z\sim2$ in the literature, $\sigma_z=0.05-0.1$ (e.g., \citealp{muzzin13,laigle16,straa16}), galaxies only with photometric redshift cannot be used for the cross-correlation analysis. 
In general, however, the majority in a given galaxy sample do not have spectroscopic redshift. 
In that sense, our results of the CCFs analysis applied to observational data are biased toward the CGM and IGM around galaxies with spec-$z$ measurements.

On the other hand, galaxies only with photo-$z$ can be used for the overdensity analysis that examines the IGM--galaxy connection on large scales (beyond tens of $h^{-1}$ Mpc). Indeed, photo-$z$ galaxies are used as a tracer of the large-scale structure (e.g., \citealp{nakata05,Blake07,Trevese07,tanaka08,hayashi19}).
However, there are still the following potential problems in using photo-$z$ galaxies for the analysis.
First is the contamination of galaxies whose real redshifts are out of the redshift range for 2D LoS data. It must always happen, even if the redshift range is wider than the mean photo-$z$ error. 
Second is a difficulty in statistically confirming a correlation when the cylinder volume is large (see also Appendix A). 
Both this study and those in the literature show the presence of IGM--galaxy connection up to $\sim 10$\,$h^{-1}$\,Mpc (e.g., \citealp{adelberger03,adelberger05,chen05,ryan-weber06,FG08,rakic11,rakic12,rudie12,font-ribera13,profX13,tejos14,bielby17}). 
In addition, according to our tests in Appendix A, we suggest that a cylinder length less than $\Delta z=0.01$ (corresponding to $9\,h^{-1}$\,Mpc) might be able to capture the large-scale structure in $\delta_\text{F}$. 
If either the cylinder length or radius is larger than the above scale, the large-scale structure traced by cosmic web and galaxies will be attenuated, and thus both $\langle \delta_\text{F} \rangle$ and $\Sigma_\text{gal}$ would become close to zero.  
In that case, the slope of the $\langle \delta_\text{F} \rangle$--$\Sigma_\text{gal}$ relation would become indistinguishable.

Based on all these arguments, the cross-correlation method is a promising method to study the variation of the IGM--galaxy connection over $1$\,$h^{-1}$\,Mpc scale using actual observational data. 
On the other hand, the overdensity analysis applied to photo-$z$ samples can probe the IGM--galaxy connection beyond several tens of $h^{-1}$\,Mpc scale.


\section{Summary}

We systematically investigate the connection between galaxies and CGM/IGM, particularly traced by Ly$\alpha$ forest absorption. In this study, we use cosmological hydrodynamic simulation (\citealp{shimizu19,Nagamine20}) and demonstrate the CGM/IGM--galaxy connection using two methods: one is the cross-correlation analysis, and the other is the overdensity analysis proposed by \citet{mukae17}. Using our simulation, we also calculate CCFs of relative gas density (both total and {\sc Hi}) around galaxies. All parameters for our analyses are chosen to match the observations presented in M21. The main results of this paper are summarized below. 

\begin{enumerate}
    \item We calculate CCFs between Ly$\alpha$ forest transmission fluctuation ($\delta_\text{F}$) and galaxies as shown in Figure~\ref{fig:IM_all_CCF}. The CCF obtained from all galaxies reproduce the one from LBGs in the literature \citep{adelberger05,bielby17}.
    Further investigations based on subsamples divided by $\Mstar$, $M_\text{DH}$, SFR, and sSFR of simulated galaxies show following trends and variations in the CCF. 
    For the $\Mstar$ and $M_\text{DH}$ subsamples, we find a clear trend that the CCF signal becomes stronger with increasing galaxy masses. 
    We also confirm that the turnover radius of CCFs becomes smaller with increasing mass (see Figure~\ref{fig:IM_all_CCF}--2), indicating  stronger clustering of massive galaxies around density peaks. Additionally, from the best-fit parameters of power-law fitting for the CCFs, clustering length $r_0$ and slope $\gamma$ are found to become longer and steeper with increasing galaxy masses in the IGM regime ($r\geq1$\,$h^{-1}$\,Mpc: see also Figure \ref{fig:fit_IM_all_CCF}).  
    For the SFR samples, we find that they tend to have stronger signals with increasing SFR. Such a trend for SFR samples is also linked to the mass dependence of CCF, because $M_\star$ and SFR is almost linearly related with each other through  star-formation main sequence of galaxies.
    However, we do not identify clear trends in the CCF of sSFR samples. 
    \item We measure CCFs between gas density fluctuation ($\delta_{\rho_\text{gas}}$ and $\delta_{\rho_\text{{\sc Hi}}}$) and galaxies in Figures \ref{fig:IM_tgas} and \ref{fig:IM_hgas}. Overall trends of CCFs are similar to those of CCFs in $\delta_\text{F}$ except for SFR--(v) and sSFR--(iii). 
    It indicates that the variation in $\delta_\text{F}$ CCF reflects different relative gas densities around galaxies; i.e., galaxies with higher mass and SFR generally reside in higher density gas, and vice versa. 
    For the SFR--(v) and sSFR--(iii) subsamples, we find the highest CCF signal at $r=0.3-2$\,$h^{-1}$\,Mpc among all SFR and sSFR samples. 
    Because the two subsamples have a mild bimodal halo mass distribution with two peaks at $M_\text{DH}\sim10^{11.3}$\,M$_\odot$ and $M_\text{DH}\sim10^{12.5}$\,M$_\odot$, their highest CCF signals are probably due to high-density regions where massive host halos reside.
    We suggest that the observed variation in the CCF is caused by the dependence of gas density (both total and {\sc Hi}) around galaxies. 
    \item Our overdensity analysis between galaxy overdensity $\Sigma_\text{gal}$ and mean IGM fluctuation $\langle \delta_\text{F} \rangle$ is presented in Figure~\ref{fig:oe}. 
    We statistically identify anti-correlations from all subsamples of $\Mstar$--11, $\Mstar$--10, $\Mstar$--9 and ALL. In addition, we also find that their slopes are decreasing with increasing $\Mstar$, although within the error. 
    It suggests that galaxies in the $\Mstar$--11 subsample are more strongly correlated with higher density gas than those in $\Mstar$--9 in terms of their spatial distribution. 
    \item Considering all of our results together, we conclude that the mass, particularly the dark matter halo mass, is the most sensitive parameter to determine the Mpc-scale gas-density environment around galaxies. 
    Galaxies in massive halos tend to be clustered in higher density regions of the cosmic web, resulting in a CCF with a higher amplitude, greater $r_0$,  steeper $\gamma$, and shallower anti-correlation between $\langle \delta_\text{F} \rangle$ and $\Sigma_\text{gal}$ at $r\geq1$~$h^{-1}$~Mpc. 
\end{enumerate}

Overall, our analyses confirm the strong connection between galaxies, dark matter halos, and IGM, providing further support for the gravitational instability paradigm of galaxy formation within the concordance $\Lambda$CDM model. 
Future observations of CCF studies between galaxies, {\HI}, and metals will provide useful information on the interaction between them and the details of feedback mechanisms which is important for the theory of galaxy formation and evolution such as galactic wind  and associated ejection of metals into IGM.  

By comparing the results of Figures~\ref{fig:IM_all_CCF} \& \ref{fig:fit_IM_all_CCF} against predictions of linear perturbation theory, we can infer the mean bias parameters of Ly$\alpha$ forest and galaxies relative to the underlying dark matter density field \citep[e.g.,][]{croft16,bielby17,kakiichi18,meyer19}. 
In addition we can perform cross-checks by computing such bias parameters directly from our simulation output, and compare with those inferred from linear theory framework. Such bias parameters will further provide additional checks against the gravitational instability paradigm, and we plan to carry out such analyses in our further work.

\acknowledgments
We are grateful to Drs. M. Rauch, A. Meiksin, H. Yajima, D. Sorini, T. Suarez Noguez, K. Kakiichi and R. A. Meyer for helpful discussions. We also appreciate the referee and the editor for providing the constructive suggestions and comments to improve our manuscript.  
RM acknowledges a Japan Society for the Promotion of Science (JSPS) Fellowship at Japan. 
KN is grateful to Volker Springel for providing the original version of {\small GADGET-3}, on which the {\small GADGET3-Osaka} code is based on. 
Our numerical simulations and analyses were carried out on the XC50 systems at the Center for Computational Astrophysics (CfCA) of the National Astronomical Observatory of Japan (NAOJ), the XC40 system at the Yukawa Institute for Theoretical Physics (YITP) in Kyoto University, and the {\small OCTOPUS} at the Cybermedia Center, Osaka University as part of the HPCI system Research Project (hp180063, hp190050).
This work is supported by the JSPS KAKENHI Grant Numbers JP18J40088 (RM), JP17H01111, 19H05810 (KN), and JP19K03924 (KS). 
KN acknowledges the travel support from the Kavli IPMU, World Premier Research Center Initiative (WPI), where part of this work was conducted. 
We acknowledge the Python programming language and its packages of numpy, matplotlib, scipy, and astropy \citep{astropy13}.

\clearpage

\appendix

To understand the impact on actual observational data, we conduct several tests for generating 2D LoS maps and the cross-correlation analysis with the actual observational data in M21 in our mind.
In this appendices we briefly show our results. Note that we only calculate the CCFs beyond $r\geq1$\,$h^{-1}$\,Mpc for those tests in order to directly compare with the observational results presented in M21.

\vspace{1cm}

\section{2D LoS maps}
\label{app:map}

\begin{figure}[ht!]
	\begin{center}
	\includegraphics[width=0.6\linewidth]{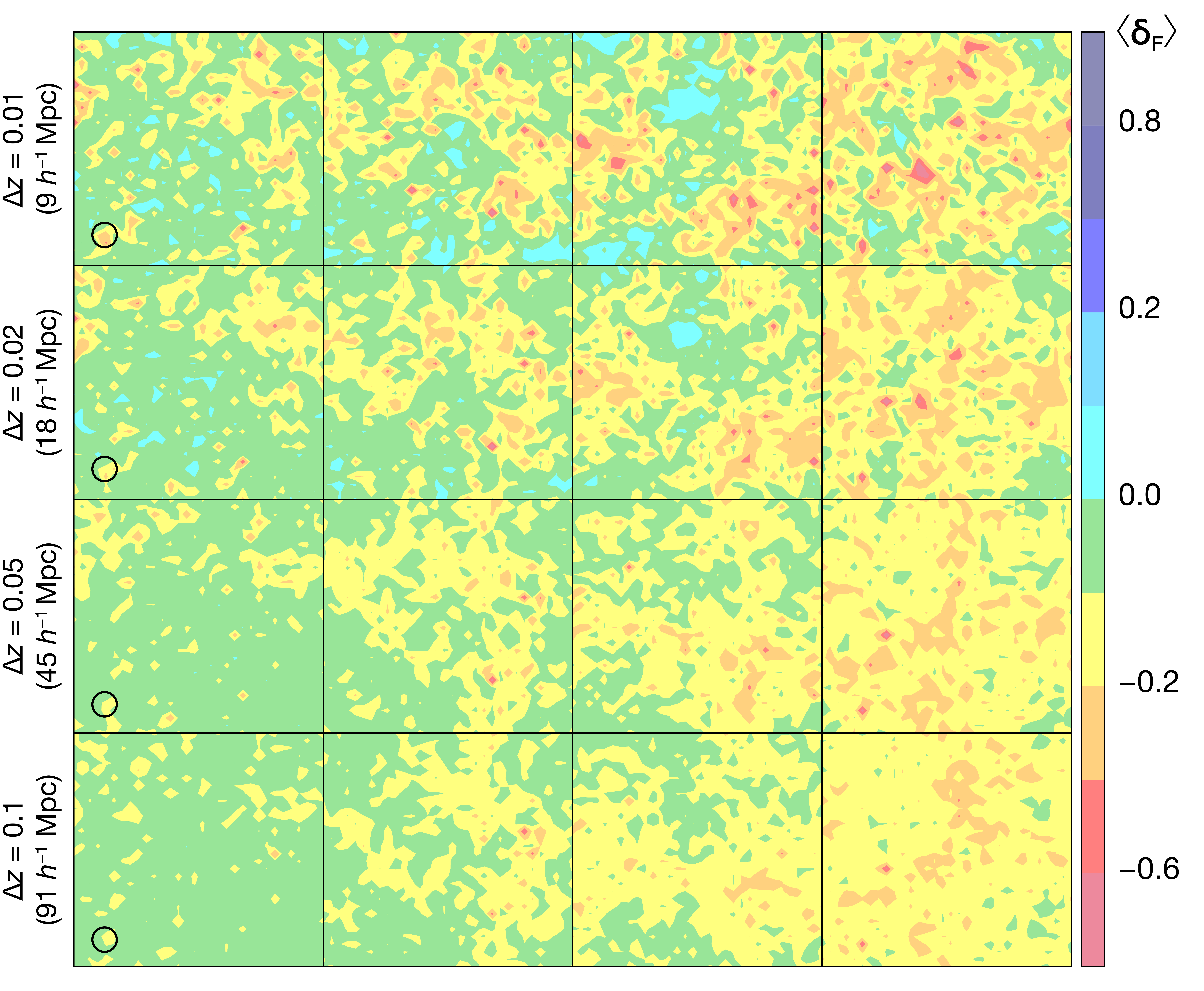} 
	\caption{
    2D LoS maps of four arbitrary redshift slices are shown in four rows. Different binning widths to generate 2D LoS maps are arranged in each column. A black circle indicates the cylinder size to estimate $\langle \delta_\text{F} \rangle$ and $\Sigma_\text{gal}$ by the overdensity analysis ($r=4.74$ $h^{-1}$ Mpc).
    }
	\label{fig:map}
	\end{center}
\end{figure}

We show 2D LoS $\delta_\text{F}$ maps binned by $9$, $18$, $45$, and $90$ $h^{-1}$ Mpc which correspond to $\Delta z=0.01$, $0.02$, $0.05$, and $0.1$ in Figure~\ref{fig:map}. Each column indicates different redshift used for the overdensity analysis in Section \ref{sec:res_oe}. 
Figure~\ref{fig:map} clearly shows that the intensity of IGM fluctuation becomes attenuated with increasing  redshift width for binning the LoS data. In addition, the dynamic range of $\langle \delta_\text{F} \rangle$ of 2D LoS maps binned by more than $45$\,$h^{-1}$\,Mpc seems to be too narrow to differentiate the environment based on galaxy properties.


\begin{figure*}
	\begin{center}
	\includegraphics[width=0.6\linewidth]{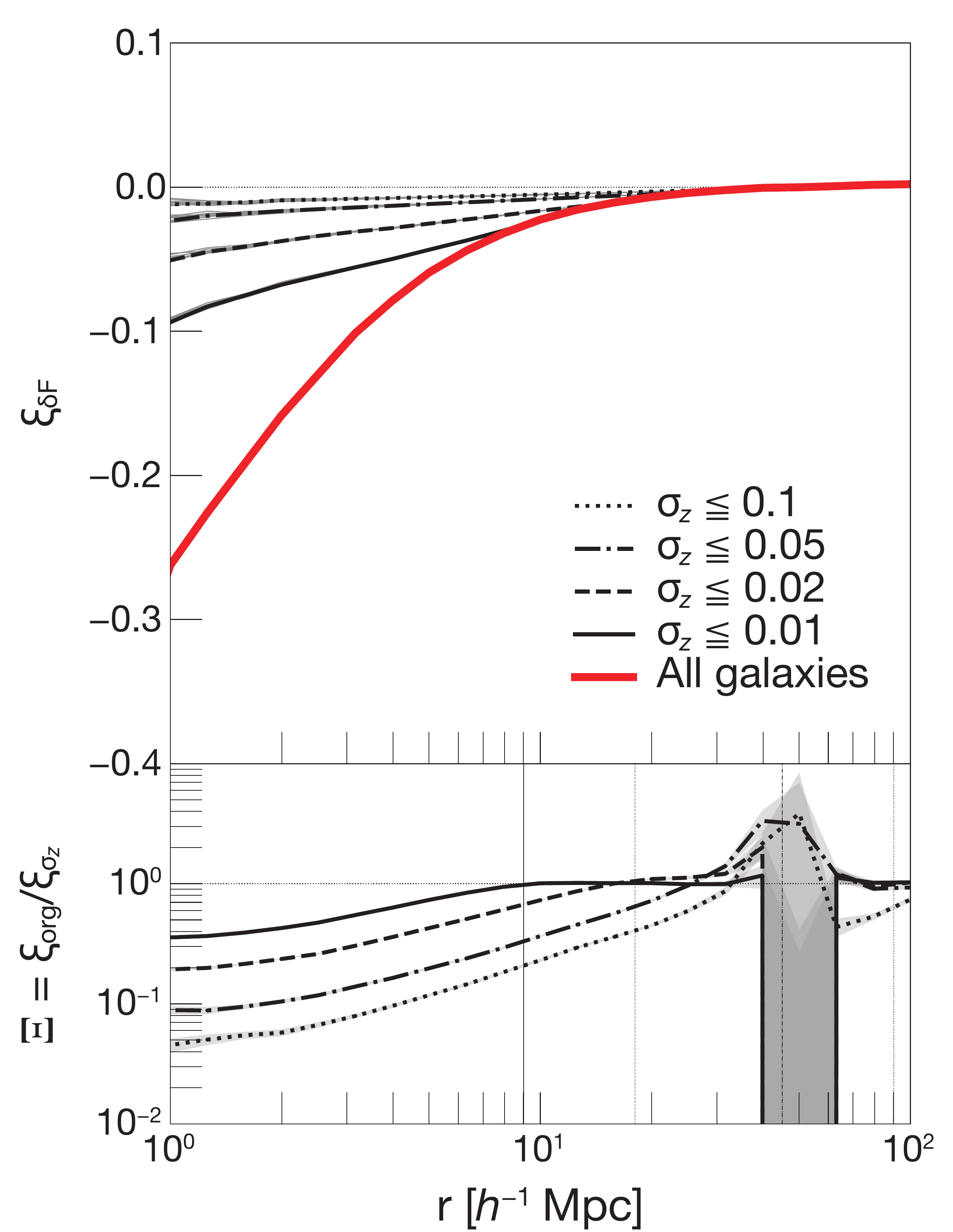} 
	\caption{
    ($Top$) The CCFs from galaxies with redshift uncertainties of $\sigma_z\leq0.1$, $0.05$, $0.02$, and $0.01$. Thin lines indicate the CCFs of $10$ tests. The dotted, dash-dotted, dashed and solid lines represent the mean of $10$ tests with $\sigma_z\leq0.1$, $0.05$, $0.02$, and $0.01$. The original CCF derived from all galaxies is colored in red. 
    ($Bottom$) The CCF ratio of original to the mean of $10$ tests ($\Xi=\xi_\text{org}/\xi_\text{$\sigma_z$}$). The gray shade shows the error of $\Xi$. Vertical four lines represent the effective radius of $\sigma_z=0.1$, $0.05$, $0.02$, and $0.01$ corresponding to $90$, $45$, $18$, and $9$ $h^{-1}$ Mpc. 
    }
	\label{fig:IM_photoz}
	\end{center}
\end{figure*}


\section{The impact of redshift measurement uncertainties on the CCF}
\label{app:photoz}

The cross-correlation analysis used in this study requires spec-$z$ sample of galaxies (see Section \ref{sec:dis_which}). Nonetheless, spectroscopic redshift measurements are not always available for galaxies in photometric images. 
Hence, here we test the usability of photo-$z$ galaxies by adding photo-$z$ errors.

We randomly add redshift uncertainties with $\sigma_z\leq0.1$ to all galaxies. Then, we recalculate the CCF using the reassigned galaxy redshift $z_\text{use}=z_\text{real} \pm \sigma_z$, where $z_\text{use}$ and $z_\text{real}$ are the redshift used to calculate the CCF and the original one in our simulation, respectively. 
This process is carried out $10$ times. The CCF of each routine and the mean of $10$ tests are shown by thin and thick black lines in Figure~\ref{fig:IM_photoz}. 
We also carry out the same tests with $\sigma_z\leq0.05$, $\sigma_z\leq0.02$ and $\sigma_z\leq0.01$. 
The actual distances corresponding to $\sigma_z$ values are ($\sigma_z=0.1$, $0.05$, $0.02$, $0.01$) = ($90$, $45$, $18$, $9$) $h^{-1}$ Mpc at $\langle z \rangle=2.3$.
In the bottom panel of Figure \ref{fig:IM_photoz}, the ratio of CCF from all galaxies colored in red ($\xi_\text{org}$) to the mean of the CCFs from galaxies in consideration of redshift uncertainties ($\xi_\text{$\sigma_z$}$)
is also presented ($\Xi=\xi_\text{org} / \xi_\text{$\sigma_z$}$).

We find that all CCF signals become weaker with respect to the original CCF colored in red, though the  scatter of CCF signal among $10$ tests is quite small.
In particular, the CCF signal becomes insignificant for the data with $\sigma_z\leq0.1$ or $0.05$. 
It suggests that galaxy data set with such a large photo-$z$ errors is useless for cross-correlation analysis. 
However, the other two cases with $\sigma_z\leq0.02$ or $0.01$ still show some signals at the center, albeit they are weak. Due to the signal detection, galaxies with $\sigma_z\leq0.02$ may be usable for calculating CCFs.

Another interesting result from this test is the radius where $\Xi$ becomes approximately one. 
Within $r<40$ $h^{-1}$ Mpc, each sample reaches $\Xi=0$ at a radius which is equivalent to the actual distance of $\sigma_z$ (see also vertical lines in the bottom panel of Figure \ref{fig:IM_photoz}). 
These results indicate that a data set with redshift uncertainties less than 1\,$h^{-1}$ Mpc (or $\sigma_z\sim0.001$) would be necessary to obtain a true CCF. 

Many high-$z$ galaxies often have photo-$z$ estimates. In the literature, such photo-$z$ uncertainties have been evaluated as $\sigma_z=(0.007-0.021) \times (1+z)$, which corresponds to $\sigma_z=0.023-0.070$ at $z=2.35$ (e.g., \citealp{laigle16,straa16}). 
Considering our tests shown in Figure \ref{fig:IM_photoz}, galaxy data set with current photo-$z$ measurements only are not useful for cross-correlation analysis. 
In order to derive an accurate CCF, galaxy samples with good spectroscopic redshifts are needed.

\section{Cosmic variance}

If a survey volume is not large enough, the cosmic variance must affect the CCFs (both amplitude and shape). 
We evaluate the effect on the CCF by limiting the volume to $\Delta z=0.1$ and $0.05$, corresponding to $91~h^{-1}$~Mpc and $45~h^{-1}$~Mpc in redshift direction. Note that because large-scale fluctuations are missed due to a limited simulation volume, we inevitably underestimate the effect of cosmic variance on large scales. 
We find that the resultant CCFs scatter around the original one obtained from all galaxies in the entire volume with a wide variation of power-law $\gamma$ and $r_0$ parameters of Equation (8). It suggests that cosmic variance also influences both the slope and clustering-length of the CCF. 
Therefore, when we compare the CCFs of two different galaxy populations, it would be desirable to match their redshift coverage.

\section{CCFs obtained from small sample size}
\label{app:number}

In Appendix~\ref{app:photoz}, we discussed the uncertainty introduced by using photo-$z$ data, and the need for more accurate spec-$z$ measurements for cross-correlation analysis. 
However, the number of galaxies with spec-$z$ measurements are limited, and therefore the derived CCF from spec-$z$ data may suffer from small sample size and differ from true signal. 
Thus, we carry out following two tests in order to verify the effect of sample size on resultant CCF.

The first test is to change the completeness of the galaxy sample. 
We calculate CCFs by randomly selecting $0.1$, $1$, and $10\%$ of galaxies from the entire sample, and repeat this procedure $10$ times. We find that the resultant CCFs scatter around the original one. Additionally, this scatter becomes smaller with the increasing fraction and negligible in the $10\%$--samples. Therefore, we argue that at least $1\%$ of the total sample must have spec-$z$ measurements to reproduce the true CCF.

Unfortunately, we do not always know the true total number of galaxies in real observations. 
Therefore as a second test, we examine the effect of using an extremely small sample of randomly selected $5$ and $10$ galaxies and repeat it $100$ times. 
We conduct this test twice by changing the galaxy selection method: one is completely random, while the other is to select only galaxies whose $\langle \delta_\text{F} \rangle$ within $1.7~h^{-1}$~Mpc in the radius are less than $-0.2$.
The resultant CCFs of both methods show a large dispersion around the original CCF, and the dispersion becomes smaller as the sample size increases from five to ten.  
However, intriguingly, the scatter among $100$ CCFs in the latter method becomes smaller than those by the former method.
It suggests that the true CCF cannot be obtained from a randomly-selected, extremely small sample. 
On the other hand, a CCF derived from a few galaxies which are located in similar IGM densities (i.e., a limiting $\langle \delta_\text{F} \rangle$), could still reflect their IGM environments.


\bibliographystyle{aasjournal}
\bibliography{./clam.bib}

\end{document}